\documentclass[12pt,preprint]{aastex}

\newcommand{\ssfr}{\mbox{$\Sigma_{\rm SFR}$}}

\newcommand{\ishape}{{\tt ishape}}
\newcommand{\tinytim}{{\tt TinyTim}}

\newcommand{\vi}{\mbox{$V\!-\!I$}}
\newcommand{\viz}{\mbox{$(V\!-\!I)_0$}}
\newcommand{\ubz}{\mbox{$(\ub)_0$}}
\newcommand{\bvz}{\mbox{$(\bv)_0$}}
\newcommand{\msun}{\mbox{${\rm M}_{\odot}$}}
\newcommand{\szcut}{\mbox{$0.4$}}
\begin{document}
\title{Hubble Space Telescope imaging of a peculiar stellar complex in NGC~6946
 \footnote{Based on observations with the NASA/ESA Hubble Space 
 Telescope.}
}
\author{S{\o}ren S. Larsen
  \affil{UC Observatories / Lick Observatory, University of California,
         Santa Cruz, CA 95064, USA}
  \email{soeren@ucolick.org}
\and
  Yuri N. Efremov
  \affil{Sternberg Astronomical Institute, MSU, Moscow 119899, Russia}
\and
  Bruce G. Elmegreen
  \affil{IBM Research Division, T.J. Watson Research Center,
    P.O. Box 218, Yorktown Heights, NY 10598, USA}
\and
  Emilio J. Alfaro
  \affil{Instituto de Astrof{\'\i}sica de Andaluc{\'\i}a (CSIC), Apdo.\ 3004,
         18080 Granada, Spain}
\and
  Paolo Battinelli
  \affil{Osservatorio Astronomico di Roma, viale del Parco Mellini 84,
         I-00136 Roma, Italy}
\and
  Paul W. Hodge
  \affil{Astronomy Department, University of Washington, Box 351580,
  Seattle, WA 98195--1580, USA}
\and
  Tom Richtler
  \affil{Grupo de Astronom\'{\i}a, Departamento de F\'{\i}sica,
         Casilla 160-C, Universidad de Concepci\'on, Concepci\'on, Chile}
}

\begin{abstract}
  The stellar populations in a stellar complex in NGC~6946 are analyzed 
on images taken with the Wide Field Planetary Camera 2 on board the Hubble 
Space Telescope. The complex is peculiar by its very high density of
stars and clusters and semicircular shape.  Its physical dimensions are about 
the same as for the local Gould Belt, but the stellar density is 1 -- 2 orders 
of magnitude higher.  In addition to an extremely luminous, $\sim15$ Myr old
cluster discussed in an earlier paper, accounting for about 17\% of the 
integrated $V$-band light, we identify 18 stellar 
clusters within the complex with luminosities similar to the brightest open 
clusters in the Milky Way.  The 
color-magnitude diagram of individual stars in the complex shows a paucity 
of red supergiants compared to model predictions in the 10--20 Myr age 
range for a uniform star formation rate. We thus find tentative evidence for 
a gap in the dispersed star formation history, with a concentration of star
formation into a young globular cluster during this gap.  Confirmation of this 
result must, however, await a better understanding of the late evolution of 
stars in the corresponding mass range ($\ga 12$ M$_{\odot}$).  
A reddening map based on individual reddenings
for 373 early-type stars is presented, showing significant variations in
the absorption across the complex. These may be responsible for some of the
arc-like structures previously identified on ground-based images. 
We finally discuss 
various formation scenarios for the complex and the star clusters within it.  
\end{abstract}


\section{Introduction}

  NGC~6946 is one of the nearest large spiral galaxies beyond the Local 
Group and has been scrutinized at nearly all wavelengths from X-rays 
\citep{sch00}, over optical and infrared studies \citep{bon88,mal96,bian00} 
to sub-millimeter \citep{tac89} and radio wavelengths \citep{bv92,eh93}. The 
galaxy is classified as an Sc--type spiral, with a strong nuclear starburst 
\citep{eng96,elm98} as well as a high level of star formation activity 
throughout the disk \citep{deg84,sau98}. NGC~6946 has hosted 6 supernovae
within the last century, another indication of vivid star formation in 
the galaxy \citep{sch94}.  Its exact distance is still somewhat 
uncertain, partly because the galaxy is located at a relatively low galactic
latitude ($b=12\deg$) and is subject to significant foreground extinction
from our Galaxy. \citet{bh84} give the reddening towards NGC~6946 as
$A_B=1.62$ while the COBE/DIRBE maps by \citet{sch98} give a somewhat lower 
value of $A_B=1.48$.  Throughout this paper we will adopt the latter value.
For conversion between $B$-band extinction and other bands we use
the reddening law by \citet{car89}.  The distance to NGC~6946 is listed 
as 5.5 Mpc in the \emph{Nearby Galaxies Catalogue} \citep{tul88}, but 
more recent values are $5.9\pm0.4$ Mpc \citep{ksh00}, based on 
blue supergiants, or $5.7\pm0.7$ Mpc \citep{ea96} based on Type II supernovae.
Here we use a distance of 5.9 Mpc to NGC~6946.

  In spite of its relative proximity and numerous studies of NGC~6946 at 
many wavelengths, a peculiar stellar complex in 
one of its spiral arms has largely escaped attention since it was first 
discovered by \citet{hod67}. Nevertheless, this object is quite conspicuous 
on optical images and was noticed by \citet{lr99} on images taken with
the 2.56 m Nordic Optical Telescope (NOT) during a search for young massive 
star clusters in nearby galaxies. The Hodge complex is host to one of the 
brightest young star clusters currently known in the disk of any spiral galaxy, 
with an estimated age of about 15 Myr and an absolute visual magnitude of
$M_V=-13.2$ \citep[hereafter Paper II]{lar01}.  It is located about 3 arc 
minutes (4.8 kpc) to the W of the galaxy center, at the end of a sub-branch 
of one of the main spiral arms.  The complex is remarkable by its circular 
outer boundary, with a diameter of about $22\arcsec$ or $\sim600$ pc.  A 
color picture from the NOT, showing the location of the complex within 
NGC~6946, was presented by \citet[hereafter Paper I]{eel00}.  In addition to 
the young ``super--star cluster'' (YSSC) they identified a number of other 
bright objects within the complex. Crowding problems and limited resolution 
on the ground-based images made it difficult to distinguish between clusters 
and individual luminous stars, but because of the high luminosities of many
of these objects, they were assumed to be mostly star clusters.  Rough age 
estimates were obtained from the $UBV$ colors, indicating ages similar to 
that of the YSSC.

  On the NGC~6946 extinction map by \citet{trew98}, the complex can be
identified as a ``cavity'' of reduced extinction, presumably because star 
formation has cleared out much of the obscuring material. However, significant 
variations in the integrated $V-I$ color across the surface of the region can 
still be seen, presumably an indication of considerable reddening variations
(Paper I). This might also explain -- at least partly -- a number of 
arc-like features noted by \citet{hod67}, who suggested that the origin 
of these arcs might be related to ``super-super nova'' explosions
\citep{shk60}, and that the apparently similar structures in the LMC
Constellation III might have originated in the same way. This possibility 
has been further discussed by \citet{efr01}.  The $7\mu$ and $15\mu$ ISOCAM 
maps by \citet{mal96} show hints of the complex, but it is not apparent on 
450$\mu$ and $850\mu$ JCMT/SCUBA data \citep{bian00}. A CO map \citep{tac89} 
of the central $6\farcm75$ of NGC~6946 shows a region of enhanced CO density, 
corresponding to the spiral arm in which the complex is located, but 
higher-resolution CO data (Walsh et al., in preparation) do not show a 
particularly high CO density in the vicinity of the complex.  \citet{kam93} 
identified a number of HI holes in the disk of NGC~6946 on high-resolution 
(15\arcsec) HI maps, possibly cleared by star formation.  Many of these HI 
holes are located in the western part of the galaxy, but none of them coincides 
with the exact location of the complex.  Thus, while much of the original 
molecular 
gas may have been consumed and/or dissociated by star formation, the 
complex does not yet appear to have caused a significant hole in the HI disk 
around it.  The NOT data also show isolated regions of H$\alpha$ emission 
near the center of the complex and along the rim, coincident with the regions 
of enhanced absorption (Paper I), indicating that some star formation may 
still be taking place in the complex or at least that it has proceeded 
until very recently.

  In this paper we analyze new images of the complex, taken with the Wide 
Field Planetary Camera 2 (WFPC2) on board the Hubble Space Telescope (HST). 
A color image of the complex and its surroundings, generated from our WFPC2 
images, is shown in Figure~\ref{fig:n6946rgb}.
In Paper II we have used these data to study the young massive cluster that 
dominates the complex and we now turn to the numerous other objects, 
including a number of other clusters and the field stars. We present a 
color-magnitude diagram for stars in the complex and analyze its
recent star formation history, probably making NGC~6946 one of the
most distant galaxies for which such an analysis has been attempted.
We also discuss possible formation scenarios.

\section{Data reduction}

  The data were obtained in Cycle 9 and consist of integrations in the
F336W ($U$), F439W ($B$), F555W ($V$) and F814W ($I$) bands with exposure
times of 3000 s, 2200 s, 600 s and 1400 s,
respectively. Shorter integrations in each band were also obtained, in
case the central pixels of the main cluster would be saturated. However,
the cluster turned out to be extended enough that this was not a problem
and hence these shorter integrations were not used. A 300 sec exposure in
F656N was also taken to trace H$\alpha$ emission.  The entire complex
is comfortably contained within the $36\arcsec\times36\arcsec$ field of
the Planetary Camera (Fig.~\ref{fig:clmap}) and we only consider data in 
the PC chip in this paper. The pixel scale of $0\farcs045$ per pixel 
corresponds to 1.29 pc at the adopted distance.

  Initial processing of the images were performed ``on-the-fly'' by the
standard calibrating pipeline at STScI. Subsequently, the YSSC was 
subtracted from the images using the ELLIPSE task in the STSDAS package, 
allowing photometry to be obtained for fainter objects near the cluster.
Objects were detected on a sum of the F555W and F814W images using the
DAOFIND task in the DAOPHOT package, running within IRAF\footnote{IRAF is
distributed by the National Optical Astronomical Observatories, which are
operated by the Association of Universities for Research in Astronomy,
Inc.\ under contract with the National Science Foundation}, using a
$3\sigma$ detection threshold. Aperture and PSF-fitting photometry was 
then obtained by the PHOT and ALLSTAR tasks in DAOPHOT, using a 
two-iteration procedure whereby additional objects were detected during
a second pass of DAOFIND on the object-subtracted image generated by
the first run of ALLSTAR \citep{ste87}. For the ALLSTAR photometry, the PSF 
was empirically determined using stars in uncrowded regions.

  The $V$ and $I$ data were far deeper than the $B$ and $U$ data, but many
of the brighter objects in the field nevertheless had adequate photometry
in all four bands. Extensive completeness tests were done for the $V$
and $I$ PSF photometry, by populating the PC image with artificial stars
having magnitudes between $V=19.0$ and $V=27.5$ at 0.5 mag intervals,
and with \vi\ colors between $-1.0$ and $+4.0$ at 1.0 mag intervals.  For
each $V,\vi$ combination, 4 completeness tests with 500 artificial stars
each were done. The stars were added at random positions within the
central $600\times600$ pixels of the PC frame, but with required separations
larger than 10 pixels.
Thus, the completeness was determined as a function of both $V$ 
magnitude and $V-I$ color, which is essential in a case like this where
the objects span a wide range in color. 

  Aperture photometry was obtained both in an $r=3$ pixels aperture
and in an $r=11$ pixels (0\farcs5) aperture.  The aperture photometry was used
mainly for extended objects, measuring colors in the smaller aperture (to 
minimize random errors) while the $r=11$ aperture was used for cluster
magnitudes to reduce the uncertainty on the aperture correction for
extended objects. The photometry was calibrated to standard $UBVI$ magnitudes
using the relations in \citet{hol95}, which refer to a $0\farcs5$ 
($r=11$ pixels) aperture.  Aperture corrections between the $r=3$ pixels 
and $r=11$ apertures (Table~\ref{tab:apc}) were determined from aperture 
photometry on synthetic images, generated by convolving the \tinytim\ PSF 
\citep{kri97} with the WFPC2 ``diffusion kernel''. For $B$, $V$ and $U$ the 
aperture corrections from $r=3$ to $r=11$ pixels apertures were essentially 
identical (to within about 0.01 mag), while the correction for the \vi\ colors
amounts to 0.06 mag.  The zero-points for 
the PSF-fitting photometry were established by comparison with the 
aperture photometry and the subsequent calibration to the standard system
was done in the same way as for the aperture photometry.

  In order to classify objects as clusters or individual stars, sizes were
measured for all objects by use of the \ishape\ algorithm \citep{lar99}. 
This algorithm estimates intrinsic object sizes by convolving an analytic 
model of the object shape with the PSF, adjusting the FWHM of the model 
until the best possible match with the observed profile is obtained. In 
this way, sizes can be reliably measured even for sources that have 
significantly smaller intrinsic sizes than the PSF. Here we have assumed
King profiles with a concentration parameter $c = $ r(tidal) / r(core) = 30
for the intrinsic object profiles and PSFs generated by \tinytim. The modeling
also includes a convolution with the WFPC2 diffusion kernel.  Alternatively,
we could have used e.g.\ the ALLSTAR sharpness estimate to identify extended
objects. A comparison of the ALLSTAR sharpness parameter and \ishape\ FWHM 
estimates for objects in the PC frame showed that the two methods yielded 
consistent results in most cases, although objects that were clearly recognized
as extended by \ishape\ would occasionally be classified as point sources by 
ALLSTAR. In such cases, visual inspection of the images and measurements of 
their FWHM values with the IMEXAMINE task in IRAF would generally support the 
\ishape\ measurements.  For a detailed description of \ishape\ and tests of 
its performance, we refer to \citet{lar99} and \citet{lfb01}.

\section{Results}
\label{sec:results}

  Figure~\ref{fig:clmap} shows the PC field of view, with the young super-star
cluster clearly visible as the brightest object in the field, near the center 
of the image. Within a radius of $r=300$ pixels from the center of the
image, the complex has a total integrated magnitude of $V=15.0$ (including
the YSSC), corresponding to $M_V=-15.0$. With $M_V=-13.2$ (Paper II), the 
cluster thus accounts for about 17\% of the total $V$-band light 
that we receive from the complex. This calculation does not account for the 
fact that significant absorption is likely to be present, obscuring some 
fraction of the light, so in practice the cluster may contribute with a 
smaller fraction of the total light from the complex.  Considering that the 
cluster has a mass around $10^6 M_{\odot}$ (with estimates ranging from 
$5\times10^5\msun$ to $(1.7\pm0.9)\times10^6\msun$, see Paper II), the whole 
complex probably has a mass of $\sim10^7\msun$. However, the uncertainty
on this number is considerable and the mass could range between a few
times $10^6$ \msun\ and well above $10^7$ \msun. 

\subsection{Separating clusters and stars}
\label{sec:clusters}

  Figure \ref{fig:V_sz} shows the intrinsic object sizes as a function of 
$V$ magnitude, measured in the $r=3$ pixels aperture. Individual stars are 
distributed in a narrow range of FWHM values $\la0.3$ pixels, while clusters 
have larger FWHM values. Reliable size measurements require a S/N $\sim30$
\citep{lar99}, corresponding to $V\sim22.5$.  Comparing object sizes measured 
on F555W and F814W images, we found a scatter of about 0.2 pixels around a 
1:1 relation for objects brighter than $V=22.5$, with no evident systematic 
differences between the two sets of measurements. This scatter is similar to 
the width of the strip defined by point sources near the $x$-axis in 
Fig.~\ref{fig:V_sz}, and is also consistent with previous estimates of the 
accuracy by which \ishape\ can measure sizes \citep[e.g.][]{lfb01}. Thus, 
we estimate 
that the size measurements are accurate to about 0.2 pixels for $V<22.5$.
Below this limit, much of the scatter in FWHM in Fig.~\ref{fig:V_sz} is 
likely due to measurement errors but for the brighter objects one can clearly 
distinguish between extended sources (likely clusters) and point sources 
(stars).  
  
  It is evident from Fig.~\ref{fig:V_sz} that some individual stars as bright 
as $M_V\approx-8.5$ are present in the field, making distinction between
stars and clusters based on luminosity alone infeasible.  We select cluster 
candidates as objects with FWHM$>$\szcut\ pixels, corresponding to an effective 
(half-light) radius larger than 0.8 pc. From the above considerations 
concerning the accuracy of our size measurements, this adopted size cut allows
us to confidently identify a list of extended objects.  We further impose a 
magnitude limit of $V=22.5$ ($M_V=-7.5$) as measured in the $r=3$ pixels 
aperture, resulting in 18 cluster candidates.  The $V=22.5$ cut-off is 
motivated partly by the need for adequate S/N for reliable size measurements, 
and partly by the fact that it becomes increasingly difficult to distinguish 
between real clusters and simple blends due to crowding at fainter magnitudes.  
The cluster candidates selected in this way are indicated in 
Figure~\ref{fig:clmap} and integrated photometry is given in 
Table~\ref{tab:cltab}, which also gives integrated colors and 
magnitudes for the whole complex measured in an $r=300$ pixels aperture and
for the YSSC (from Paper II).  The bright object near cluster \#2848 is 
unresolved on the HST image and is probably a Galactic foreground star 
(Paper I). Most of the 18 objects selected according to the above criteria 
are probably genuine clusters, although the census may not be complete.
It is clear from Fig.~\ref{fig:V_sz} that a fainter magnitude limit 
and/or smaller size cut would lead to more cluster candidates, but
we have chosen relatively conservative size/magnitude cuts to reduce the
risk of including blends of stars in crowded regions. Also, our selection
criteria do not include more extended groupings of stars (such as the
fuzzy object next to cluster \#2284), but these are unlikely to be bound 
clusters.

  The $V$ band magnitudes in Table~\ref{tab:cltab} are measured in the 
$r=11$ pixels aperture, making them less sensitive to size-dependent aperture 
corrections.  11 of the clusters have $M_V$ brighter than $-9$ and two objects 
are brighter than $M_V=-10$.  Although they appear relatively inconspicuous in 
comparison with the YSSC, these clusters are actually comparable 
in luminosity to the brightest open clusters in the Milky Way.  Most of them 
were also identified in Paper I, but some of the fainter sources identified
in that paper are probably individual luminous stars or chance alignments of 
several stars, which were unresolved on the ground-based images.

  A $(\bv,\ub)_0$ two-color diagram for objects brighter than $V=22.5$ is 
shown in Figure \ref{fig:bv_ub}. Clusters are shown with $+$ markers while 
stars are shown with diamond ($\diamond$) symbols.
The plot also includes two solar-metallicity stellar isochrones for ages 
of 4 Myr and 10 Myr 
from \citet{gir00} and the \citet{gir95} `S'-sequence, representing the
average colors of LMC clusters. The arrow indicates the effect of
a reddening of 0.5 in $E(\bv)$. The S-sequence is basically an age sequence,
with ages increasing along the sequence from blue to red colors. Stars and
clusters clearly have different color distributions, with most of the
clusters scattering along the S-sequence while stars occupy a wider range
in colors, giving confidence to the selection criteria.  Although the cluster 
candidates are mostly distributed along
the S-sequence, the direction of scatter is also nearly parallel to the 
reddening vector.  Thus, the scatter in cluster colors can be caused by age 
differences, by differences in the absorption, or a combination of the two. 
An additional source of scatter in the cluster colors comes from stochastic 
color variations, because much of the integrated light is 
dominated by a small number of very luminous stars \citep{gir95}. The latter
effect becomes more severe for the fainter clusters.

  Another potential age indicator is the reddening-free $Q_1$ parameter, 
$Q_1 = (\ub) - 0.69 (\bv) - 0.05 (\bv)^2$ \citep{car89,har97}. 
Figure~\ref{fig:qplot} shows model predictions for the evolution of $Q_1$ as 
a function of age, using population synthesis models from \citet{bc01}. The 
models are shown for three different metallicities, $Z=0.02$ (solar), 
$Z=0.008$ and $Z=0.004$.  From Table~\ref{tab:cltab}, most clusters have
$Q_1$ values between $-0.9$ and $-0.6$. By comparison with 
Figure~\ref{fig:qplot}, it is clear that the clusters are all young objects,
but the $Q_1$ index varies in an essentially random manner as a function of
age below log(age)$\approx$7.2, with a strong dependence on metallicity as
well. Thus, the cluster ages are not well constrained.

\subsection{Constructing a reddening map}
\label{sec:redmap}

  A reddening map of the Hodge complex was presented in Paper I, based
on its integrated \vi\ colors. Although that map showed
evidence for significant reddening variations, it
relied on the assumption that color variations were due to differential
reddening of a background illumination of constant color.  With the HST data, 
this analysis can now be refined by obtaining reddenings for individual 
early-type stars. 

  In order to determine reddenings for individual stars, we again use
the $Q_1$ parameter. For $Q_1 < 0$, the relation between $Q_1$ and 
\bvz\ can be approximated as $\bvz = 0.32 \, Q_1 - 0.024$ \citep[read off
Fig.2 in][]{har97}. Our adopted reddening law is slightly different from
the one used by \citet{har97}, but comparison with stellar models from
\citet{gir00} shows that this hardly affects the relation between
$Q_1$ and \bvz .  Thus, by comparing the observed \bv\ color of a star 
with its intrinsic color \bvz\ determined from the reddening-free $Q_1$, the 
color excess $E(\bv ) = (\bv) - \bvz$ can be established, and the $B$-band 
absorption is then $A_B = 4.0 \, E(\bv)$.

  In this way, individual $A_B$ values were determined by using the PSF 
photometry for all point sources with errors in \ub\ and \bv\ less than 0.3
mag, $\ubz < 0$ (only corrected for foreground extinction) and $Q_1<0$.
A total of 373 stars satisfied these criteria. The \bv\ and \ub\ error limits
of 0.3 correspond to an error of 0.38 in $E(B-V)$ or about 1.5 magnitudes
in $A_B$ for any individual star, which is on the same order of magnitude
as the total scatter in the observed reddenings. Even though the 0.3 mag
are an upper error limit and some stars will have smaller errors than this,
it is necessary to obtain the reddening estimate at any given point in the 
image as an average of several individual measurements. We generated our 
reddening map by convolution of the individual measurements with an 
Epanechnikov kernel with a smoothing radius of 30 pixels. Since most of the 
stars are contained within a radius of $\sim200$ pixels from the center, this 
means that an average of about 10 individual measurements is used at each 
point in the map, making it possible to detect variations in $A_B$ at about
the 0.5 mag level. Narrowing down the error limits leads to smaller errors
on any individual $A_B$ value, but also decreases the number of sampling
points.

  The resulting reddening map is shown in Figure~\ref{fig:abmap}. 
Red and blue colors indicate high and low reddening, respectively. The 
reddening scale is indicated by the wedge at the bottom of the figure, which 
spans a range between $A_B=1.5$ and $A_B=3.5$ in total extinction, i.e.\ the 
map shows a range of $A_B=0.0-2.0$ in \emph{internal} $B$-band absorption.  
Compared to the Paper I reddening map, 
Figure~\ref{fig:abmap} shows the same main features: 
Areas of increased reddening along the outer arc and near the dark clouds north 
(left) of the bright cluster, and lower absorption to the southwest (above 
and to the right). Most features outside the main complex, where the stellar
density is low, are probably just artifacts. Roughly speaking, whenever the
circular shape of the convolution kernel is present, this means that the
reddening value within that region of the map is based on only one star,
and should therefore be interpreted with caution.
  
  As a test of the reality of the features seen in Figure~\ref{fig:abmap},
we generated similar maps, but using \vi\ colors instead of \bv\ colors
to get the color excess. Another test was done by randomly dividing the stars
into two samples and generating reddening maps for each sample.  All the 
main features were consistently reproduced by the various methods: Increased 
reddening along the outer arc and near the dark clouds, and decreased 
reddening to the southwest.  The agreement with the Paper I reddening map 
lends further support to the idea that many of the apparent structures in 
the complex are (at least partially) due to obscuration.  In this respect, this 
complex may be quite different in nature from the peculiar arcs in e.g.\ the 
LMC Constellation III, where absorption probably plays only a minor role.

\subsection{Modeling of the color-magnitude diagram}

  As discussed in Section~\ref{sec:clusters}, ages of stellar clusters
are uncertain because of stochastic color variations, variable reddening 
etc. The small number of clusters, age fading and dynamical evolution 
represent further difficulties when using clusters as tracers of the star 
formation history (SFH). However, information about the SFH can also be 
inferred from the numerous \emph{field stars} that are present in the complex. 

  Figure~\ref{fig:cmd_iso} shows a $V,\viz$ color-magnitude diagram (CMD) 
for PSF photometry of point sources, with the 50\% and 25\% completeness 
limits indicated.  The blue part of the CMD is dominated by a combination
of main-sequence (MS) stars and blue supergiants as bright as $M_V\sim-9$,
while red supergiants appear at $M_V\approx-6$ with most of them being
fainter than $M_V=-5$.  The plot also includes solar-metallicity 
stellar isochrones from the Padua group \citep{ber94,gir00} for ages of 4, 
10, 16, 25, 40 and 63 Myr. It is clear that the entire CMD cannot be 
adequately represented by a single isochrone and the red supergiants
(RSGs) appear to be significantly older than the most luminous blue
stars, indicating an age spread within the complex. 

  Rough estimates of the expected contamination from Galactic stars 
can be obtained from \citet{rb85}. They do not
list data for NGC~6946 specifically, but we can use their calculations for 
NGC~2808 which is located at $b=-11\deg$ and $l=282\deg$.  As their model 
assumes symmetry around the Galactic plane and with respect to $l=0\deg$, 
this is equivalent to a position about $20\deg$ closer to the Galactic
center than NGC~6946 and slightly closer to the plane, i.e.\ this field
should slightly overestimate the contamination. For $\bv<1.3$, \citet{rb85}
predict less than 1 star with $21<V<27$ within the PC field. For $\bv>1.3$,
8 stars with $25<V<27$, 5 stars with $23<V<25$ and 2 stars with $21<V<23$
are predicted. Using Padua models and correcting for foreground extinction,
the $\bv=1.3$ limit corresponds to $\vi\sim1.3$. Counting the number 
of stars with $\vi>1.3$ in our CMD, we find that the expected contamination
rate is $\sim1\%$ for $V$ fainter than 23 and $\sim3\%$ for $V$ brighter
than 23. Clearly, contamination by foreground stars is negligible.

  Figure~\ref{fig:spat} illustrates the spatial distributions of early-type
stars brighter than $V=23.5$ ($M_V=-6.5$) (left) and for RSGs (right).  
Here we have selected ``early-type'' stars as stars with $\vi < 1.0$ and
RSGs as stars with $\vi > 1.5$.
As this division is essentially one by age, the two maps can be used to 
compare the distributions of the youngest and slightly older stars in the 
complex.  Both maps show a concentration towards the center of the complex, 
indicating that both populations are indeed associated with the complex. The 
two dark clouds near the center can be clearly seen as ``voids'' in the spatial 
distribution of blue stars. Generally, the outer boundary of the region 
occupied by these stars is fairly well defined. For the RSGs, the spatial 
distribution is more diffuse, as expected for an older population that has 
had more time to migrate away from the original birth locations.

  The maps in Fig.~\ref{fig:spat} may be affected by the distribution of 
external extinction, sampling and completeness of the photometry. In order to
reduce these effects, we also generated density maps showing the relative 
densities of blue and red stars, normalized to the total number of stars at 
each position (Figure~\ref{fig:densipop}). Again, these maps show a 
concentration of both red and blue stars towards the center of the complex, 
although the density maximum of the red population has now shifted toward the
higher $X$-values. This may, however, be an artifact of the normalization
procedure, since the number of blue stars is lower there.

In the following we 
will attempt to gain some further insight into the recent star 
formation history of the complex by reconstructing its color-magnitude 
diagram, using the Padua set of isochrones. The goal is to find a 
combination of isochrones which, when combined, provides a match to the 
observed color-magnitude diagram, and then infer the star formation 
history (SFH) from the relative weights of the isochrones.

\subsubsection{Technique}

  The basic idea behind such modeling is that the relative fractions of
stars of different ages vary from one part of the color-magnitude diagram
to another. Thus, by counting the number of stars in various parts of the 
color-magnitude diagram and comparing with model predictions, the 
combination of stellar isochrones which best fits the observed number
counts can be determined. Some early attempts to model the star formation
history of CCD fields in the LMC in this way were done by \citet{ber92}, 
who simply compared the relative numbers of red clump and main sequence 
stars, enabling them to constrain the relative SFR for young stars (dominating 
the upper main sequence) and $\sim 4$ Gyr old stars (dominating the red 
clump).  More sophisticated techniques have since been employed for the LMC 
\citep[e.g.][]{hol99,ols99,dir00,dol00,har01} and other Local Group galaxies 
\citep{gal99,wyd01}, dividing the CMD into finer grids and/or utilizing 
photometry in multiple bands.  As discussed in these papers, great care must
be taken when interpreting the results from such modeling, since different 
assumptions about reddening, distance, metallicity, stellar IMF, binary
fraction and other factors can lead to significantly different results. 
Furthermore, the models of massive stars are still very uncertain and have
well-documented problems in matching the observed CMDs of luminous stars
in the Milky Way as well as the LMC \citep[e.g.][]{chi98}.  With these 
precautions in mind, we nevertheless proceed to attempt a reconstruction of 
the CMD for the complex in NGC~6946.
  
  The isochrones were downloaded from the WWW site of the Padua group and 
are based on \citet{gir00} stellar models for masses up to 8 \msun\ and on 
\citet{ber94} models for more massive stars. For each individual isochrone
a set of synthetic ($V,I$) datapoints was generated, corresponding to
masses between 100\msun\ and a lower limit of 4\msun , below the observational 
cut-off. Unless otherwise noted, the masses were distributed according to
a \citet{salp55} stellar IMF. The $V$ and $I$ magnitudes for a given mass and
age were obtained by interpolation in the isochrone and adding an optional
offset to simulate reddening. Datapoints corresponding to masses above the 
endpoints of stellar evolution were simply assigned a luminosity of 0, thereby 
not contributing to the CMD, but still being counted for normalization 
purposes.  Photometric errors were then added to each datapoint by adding 
random offsets to the $V$ and $I$ magnitudes, scaled to the photometric 
errors of the closest star in the observed CMD. Finally, the synthetic data
were de-populated according to the completeness function determined for the
observations. Because of brighter completeness limits in $B$ and $U$,
we only used the $V$ and $I$ data for this modeling.

  For the purpose of evaluating the ``goodness'' of the fit to the observed
CMD provided by a modeled CMD we chose the $\chi^2$ statistic,
comparing the number of stars in the synthetic and observed CMDs within
a number of ``$\chi^2$ boxes''.  We compared the number of stars
observed in the $i$th $\chi^2$ box (${\rm Cnt}_{i, {\rm data}}$) 
with the density of synthetic datapoints in the same box
(${\rm Cnt}_{i, {\rm synt}}$), obtained by summing the contributions from 
each isochrone (${\rm Cnt}_{ij, {\rm synt}}$).  The best fit was then 
defined as the one minimizing the function
\begin{equation}
  \chi^2(W) = \frac{1}{N_{\rm bx}} \sum_{i=1}^{N_{\rm bx}} \frac{ \left(
         {\rm Cnt}_{i, {\rm data}}
      - c \, \sum_{j=1}^{N_{\rm iso}} W_j {\rm Cnt}_{ij, {\rm synt}} \right)^2}
         { {\rm Cnt}_{i, {\rm data}} + 1} 
  \label{eq:chsq}
\end{equation}
where $N_{\rm bx}$ is the number of $\chi^2$ boxes, $N_{\rm iso}$ is the
number of isochrones, $c$ is a normalization constant and $W_j$ are the 
weights of the isochrones. The addition of 1 to the denominator is just
a convenient way to avoid division by zero for empty boxes.

  Figure~\ref{fig:isofig} illustrates the first steps in this process: each 
of the six panels in the figure shows the CMD for a single isochrone after 
addition of photometric errors.  None of the panels in Fig.~\ref{fig:isofig} 
provides a match to the observed color-magnitude diagram, which contains
a combination of very luminous blue stars only present in the youngest
isochrones, as well as red supergiants belonging to a somewhat older
($\sim25$ Myr) population.  Each of the plots in Figure~\ref{fig:isofig} 
contains 1000 stars, but the synthetic CMDs used for the fits were populated 
with many more stars in order to eliminate the effects of Poisson statistics
in the synthetic data. Specifically, about $3\times10^5$ synthetic datapoints
were generated for \emph{each} isochrone.

  Figure \ref{fig:synt_cnst_p} shows a synthetic color-magnitude diagram
for a \emph{constant} star formation rate from 4 to 100 Myr. The upper
age limit is not important; above $\sim50$ Myr only few stars are brighter 
than the completeness limit. This synthetic CMD is already a much better
match to the observed one, although some important differences remain. 
There are now too few luminous early-type stars and too many RSGs brighter 
than $M_V\sim-5$. By comparison with Fig.~\ref{fig:cmd_iso}, this suggests 
that the youngest isochrones should be weighted more strongly, and isochrones 
at intermediate ages (10--20 Myr) should carry less weight.

  The Padua isochrones are tabulated at intervals of 0.05 in log(age), 
making Eq.~(\ref{eq:chsq}) a function of 29 variables for a range of 
6.6--8.0 in log(age).  In practice, this was a prohibitively large number
of free parameters to fit at one time, so to make the problem more
manageable we divided the isochrones into a number of groups,
fitting all isochrones within a group together. The separations between
the groups were set at log(age) = 6.7, 6.85, 7.0, 7.15, 7.3, 7.5, and 7.7. 
The $\chi^2$ minimization was done with the \emph{downhill simplex} algorithm 
as described in \emph{Numerical Recipes} \citep{press92}. Because the 
algorithm has a known tendency to get stuck at local minima, we followed
the recommended procedure of restarting it once it claimed to have
found a minimum. A stable minimum would typically be reached after 
3--4 restarts. As an additional safety precaution against reaching only
a local minimum, each simulation involved running the algorithm 5 times 
with different initial guesses for the fitted parameters (a constant
SFR, constant weights, and three randomly selected combinations of weights
for each group). In general, the final result did not depend strongly
on the initial guesses.

  Our algorithm allowed the $\chi^2$ boxes to be user-defined. Ideally, one
would hope to obtain a perfect fit between the simulated and observed data
at any point in the CMD, but in reality this is not possible. Depending on
how the boxes were defined, we could regulate the sensitivity of the results 
to various effects: For example, very broad boxes would reduce the influence 
of reddening effects.  Also, a cruder division was used to fit the
red supergiants, since the colors and relative numbers of these are more
uncertain.  In practice, the boxes were chosen in a 
trial-and-error manner, by adjusting the number of boxes and the 
color/magnitude range covered by each box until the algorithm succeeded 
in producing a synthetic CMD that resembled the observed CMD as closely
as possible.

\subsubsection{Star formation history}

  The best-fitting synthetic CMDs are shown in Figure~\ref{fig:syntfig}.
The upper-left panel shows the actual data and the $\chi^2$ boxes, while
the remaining three panels show synthetic CMDs for 
different assumptions about the internal reddening: No reddening 
($A_B=0$) and random reddenings in the intervals $0<A_B<1$ and $0<A_B<2$.  
Details about the fits are listed in Table~\ref{tab:ftab}.
Each plot contains the same number of stars as the observed CMD.  The $A_B=0$
synthetic CMD clearly provides a poor match to the observations and is 
unable to reproduce the broad color distribution of the early-type stars, 
no matter how the star formation history is fine-tuned. Visually, the
$0<A_B<2$ plot appears to provide the best match, but the $\chi^2$ is
actually lower for the $0<A_B<1$ plot. In any case, these simulations
again show that significant reddening is present in the complex.

  Instead of assigning random reddenings to the model CMDs, we could also
have used the reddening map (Figure~\ref{fig:abmap}) to de-redden the data. 
However, a color-magnitude diagram for photometry de-reddened by means of the 
reddening map did not appear significantly different from a CMD where a 
uniform reddening was assumed, presumably because the reddening map only gives 
an average reddening at each point and does not contain information about the 
scatter at any given position. We thus decided to follow the simpler approach 
of assigning random reddenings to the simulated CMDs.
  
  Figure~\ref{fig:sfh} shows the star formation histories corresponding 
to the three synthetic CMDs in Figure~\ref{fig:syntfig}. The relative
star formation rate at each age is simply the weight $W_j$ divided by
the spacing in age between the isochrones at that age.  Each curve has
been normalized to give a total stellar mass of $10^7$ \msun\ over the
past 50 Myr. The YSSC is represented by the Gaussian curve at 15 Myr,
assuming a cluster mass of $10^6$ \msun. The curve has a FWHM of 2 Myr,
corresponding to the cluster crossing time for a diameter of 20 pc and 
velocity dispersion 10 km/s \citep{elm00}.

  Although the details clearly depend on the reddening assumption, all 
three SFHs show the same qualitive features: A recent burst of star 
formation, starting at about 6 Myr and continuing up until 
the lower age limit of the isochrones, a period of essentially no star 
formation between 6 Myr and 15--20 Myr (except for the YSSC), and an older 
burst. Beyond 30 Myr or so the results become very uncertain, as even 
the brightest stars are near the detection limit. 

\subsubsection{Spatial variations in the SFH}

  In an attempt to detect any differences in the SFH between the inner and
outer parts of the complex, which might provide hints to its origin, we
performed the CMD modeling analysis for the central 100 pixels and for
$100<r<300$ pixels separately (see Figure~\ref{fig:clmap}). The observed and 
best-fitting synthetic CMDs for the two zones are shown in 
Figure~\ref{fig:syntfig_io} and the corresponding star formation histories 
are plotted in Figure~\ref{fig:sfh_io}. The normalizations of the respective 
SFHs are arbitrary. Figure~\ref{fig:sfh_io} suggests slightly different SFHs 
for the two zones: The recent burst seems to be somewhat stronger in the 
central parts, and the intermediate quiescent period may have been briefer 
here. The older burst overlaps with the age of the young super-star
cluster, and it seems likely that the cluster formed near the end of this 
initial burst of star formation. Although we have argued above that the 
results are very uncertain for ages older than $\sim30$ Myr, the smaller 
relative SFR at ages $\ga$ 30 Myr in the central zone is compatible with 
the oldest stars belonging to a general field population whose relative 
contribution is smaller in the more densely populated central parts.
However, to first order the entire complex appears to have had a quite 
uniform star formation history.

\subsubsection{Uncertainties}
\label{sec:unc}

  As already mentioned, the results can be quite sensitive to the various 
input assumptions. A shallower IMF slope, for example, leads to a weaker 6 
Myr burst, because a star forming event of any given strength produces larger 
numbers of high-mass stars.  We also tried varying the distance modulus 
by 0.5 mag to either side of the assumed value (28.4 and 29.4) but found
no strong effect on the results. Again, the best solution would be obtained
for a star formation history with an initial burst, followed by 
a quiescent period and a second burst. 

  The lower age limit of the Padua isochrones is 4 Myr, effectively
truncating our simulated SFHs at ages below 4 Myr or about half the 
estimated age of the onset of the second burst.  If star formation 
is continuing until the present day in the complex then even younger stars 
would also contribute to the CMD, and the average relative SFR of
the youngest burst would be correspondingly lower.

  A more critical ingredient in the modeling is the stellar isochrones.  
The Padua group also provides lower-metallicity models ($Z=0.008$ and 
$Z=0.004$), but for these lower metallicities the red supergiants are too 
blue to be compatible with the observed CMD. Thus, the only isochrones 
that allow us to obtain an acceptable fit to the data are the 
solar-metallicity ones.  The abundances of various elements in HII regions 
in NGC~6946 have been studied by \citet{mc85} and \citet{fer98}. The galaxy 
shows a clear abundance gradient as a function of distance from the center, 
but at the distance of the stellar complex discussed in this paper both 
studies find Oxygen abundances very similar to solar. However, no information 
is available about iron abundances.

  According to the models, the main sequence turn-off 
of a 10 Myr old stellar population corresponds to a stellar mass of about 18 
\msun.  For an age of 25 Myr the turn-off is at $\sim$ 9 \msun. The CMD is 
therefore completely dominated by high-mass stars, for which stellar models
are still plagued by uncertainties \citep{chi98}.  In particular, one 
long-standing problem has been to reproduce the observed number ratios of blue 
and red supergiant stars in young stellar populations 
\citep[e.g.][]{sal99,mm00}.  This ratio, as well as the temperature (and thus 
color) of the RSGs, depends strongly on a number of factors such as 
metallicity, stellar rotation, convection and mass-loss rates.
Isochrones with a smaller number of RSGs in the 
10 -- 20 Myr range might be able to reproduce the observed CMD in the NGC~6946 
complex without requiring an age gap. 

  A comprehensive library of stellar models and isochrones is also available
from the Geneva group \citep{ls01}. Comparison with these models provides
some useful insight into the differences between stellar models.  In 
Fig.~\ref{fig:isocmp} we compare solar-metallicity isochrones from the two 
groups for three different ages (8, 13 and 40 Myr), showing the Padua 
isochrones with solid lines and Geneva isochrones with dashed lines. As
recommended by \citet{ls01}, we have used their ``extended'' grids which 
incorporate high mass loss rates for the most massive stars. One shortcoming
of the Geneva models is that they do not list colors and magnitudes for
Wolf-Rayet stars, due to the lack of accurate atmosphere models for these
stars.  

  While the 8 Myr isochrones show good agreement between Geneva and
Padua models, significant differences exist for the older 
isochrones in the sense that the Geneva models produce cooler 
and fainter (in $V$-band) red supergiants for a given age. Another 
difference, which is 
not clearly seen from the isochrones but has important consequences for the 
CMD modeling, is that Geneva models spend more time as blue supergiants 
than the Padua models.
  Figure~\ref{fig:synt_cnst_g} shows constant-SFR simulated CMDs based
on the Geneva isochrones. The left panel is for solar metallicity ($Z=0.020$)
and the right panel is for $Z=0.008$. Compared to Fig.~\ref{fig:synt_cnst_p},
the red supergiants are about 0.5 mag redder in \vi\ for $Z=0.020$, while
the $Z=0.008$ Geneva models produce RSGs of the about the same color as
the Padua $Z=0.019$ models. Also note that Geneva models produce no
RSGs brighter than $M_V=-6$ for solar metallicity, in better agreement
with the observed CMD.  The blue part of the CMD is actually dominated
by blue supergiants older than about 15 Myr, rather than by very young
main sequence stars.
  
  When trying to reconstruct the observed CMD using the Geneva models,
we found that no satisfactory fit could be obtained for the solar metallicity 
models, primarily because of the too red color of the RSGs.  However, for 
the best-fitting solar metallicity models, two bursts were still preferred, 
with the youngest burst occurring at $\sim 6$ Myr and the older burst at 
10--15 Myr. This difference in the age of the older burst compared to the 
Padua model fits is due to the fainter RSGs in the Geneva models at a given 
age. The $Z=0.008$ Geneva models produce a somewhat better 
fit to the observed CMD and the resulting SFH is very similar to the one 
based on the Padua models. A characteristic problem when using the Geneva 
models is that the large number of blue supergiants produced by populations
older than about 15 Myr cannot easily be reconciled with the need for 
a very young population, because the blue part of the CMD tends to be too 
crowded below $M_V\approx-6$ in the presence of both. 

  Color-magnitude diagrams for other nearby galaxies do show some red 
supergiant stars in the magnitude interval predicted by the models for 
ages of 10--20 Myr, i.e.\ in the range $-8 < M_V < -5$ 
\citep{wil90,lyn98,col99}.  However, most of these CMDs are for 
lower-metallicity environments and thus may not be directly
comparable to the situation in NGC~6946. \citet{mas98} finds a steady
decrease in the luminosity of the brightest red supergiants for selected
fields in NGC~6822, M33 and M31 as a function of metallicity, with the 
brightest RSGs in M31 having $M_V\approx-6$. This is only slightly
brighter than the brightest RSGs in the NGC~6946 complex. 

  Another shortcoming in our modeling is that we do not treat binary 
stars. For equal-mass binaries this would effectively make the integrated 
magnitudes brighter by 0.75 mag, while the situation is more complicated 
if the components have different masses. However, there is no conceivable 
way in which binaries could account for the ``missing'' red supergiants, 
except perhaps for their impact on stellar evolution in close systems.

\subsection{Cluster ages revisited}

  Although the reddening presumably varies substantially at any given position
within the complex, the reddening map (Fig.~\ref{fig:abmap}) provides some 
help in breaking the age-reddening degeneracy for stellar clusters. We thus
return to the problem of determining cluster ages, now taking advantage
of the improved knowledge about reddening. Table~\ref{tab:clages} lists
data for each of the 18 clusters. Internal $A_B$ values from the 
reddening map are listed in the second column of the table, and the last two
columns give the S-sequence ages derived from $UBV$ colors uncorrected and 
corrected for internal reddening, respectively. The reddening correction
does affect the ages somewhat, but not dramatically.  \citet{gir95} quote an
internal scatter of 0.14 in the S-sequence ages. In addition to this, 
photometric errors typically contribute with 0.2 to 0.4 in the uncertainty
on log(age).  Also, many of the same uncertainties concerning the evolution 
of massive stars dicussed in Section~\ref{sec:unc} affect estimates of 
cluster ages from their integrated colors.  

  A histogram of the cluster ages in the last column of Table~\ref{tab:clages} 
is shown in Figure~\ref{fig:clages}. Note that the YSSC (age = 15 Myr) is 
not included in this plot.  Like the SFH plots based on individual stars in 
Fig.~\ref{fig:sfh}, Figure~\ref{fig:clages} hints at two episodes of 
cluster formation, with a quiescent period around $\sim$15--25 Myr.
Formally, the age distribution of the youngest clusters appears to peak 
earlier than the onset of the youngest burst of field star formation,
but considering the uncertainties in the age determinations for clusters
as well as stars it is not clear that this difference is significant.
Also, the relative numbers of clusters of different ages are not easily
comparable.  Between an age of 6 Myr and 20 Myr a stellar cluster fades by 
$\approx 1.5$ mag in $V$ \citep{bc01}, introducing a strong bias towards the 
youngest objects in a luminosity-limited sample, and dynamical destruction 
processes further reduce the number of clusters observed after $\sim20$ Myr. 
Thus, it is quite likely that cluster formation took place throughout the 
star forming history of the complex, but most of the older clusters have 
faded beyond our detection limit or dissolved by now.  Also note that the 
two oldest clusters, \#99 and \#2848, are located outside the main body of 
the complex and may be unrelated field objects.

\section{Discussion}

  How does the complex in NGC~6946 compare with other large-scale star 
forming structures? In our Galaxy, the Sun itself is situated within a large 
structure, known as the Gould belt. The Gould belt has an estimated diameter 
of about 600--1000 pc, very similar to that of the complex in NGC 6946. 
The mean age is about 30 Myr, but with a large age spread \citep{sf74,pop97}. 
From our vantage point within the Gould belt it is difficult to measure its 
total mass, but estimates range between $5\times10^4$ M$_{\odot}$ 
\citep{com94} and $5\times10^5$ M$_{\odot}$ \citep{tay87}. \citet{com01} 
quotes a total integrated $B$ magnitude of $M_B\approx-12.7$ for the Gould 
Belt. From our F439W images we get an integrated magnitude of $M_B=-15.1$ 
for the NGC~6946 complex (Table~\ref{tab:cltab}), i.e.\ about a factor of 10
brighter than the Gould Belt. Alternatively, our crude estimate of the total 
stellar mass (Section~\ref{sec:results}) of $\sim10^7\msun$ is between 20 
and 200 times higher than the above estimates for the Gould belt.  The total 
ages are not well known for either the Gould belt or the NGC 6946 complex, 
but adopting characteristic ages of 30 Myr and 15 Myr, respectively, the 
mean SFR in the NGC 6946 complex appears to have been some 20 -- 400 times 
higher than in the Gould belt.

  Isolated arc-shaped or arc-including stellar complexes have been noted 
in a number of other galaxies \citep{efr01a}, but detailed information
about ages, masses and integrated luminosities is generally lacking. As
the highest level in the hierarchy of clustering of young stellar 
populations, stellar complexes are omnipresent \citep{efr95,elm00a},
but isolated complexes similar to the one investigated here are rare.
For a complex in M83, \citet{com01} estimated $M_B=-11.4$, a diameter of 450 pc 
and an age of $\sim10^7$ years. Thus, the M83 complex is quite comparable in 
total luminosity and linear extent to the Gould belt, but about 30 times 
fainter than the complex in NGC 6946.  Since it is probably also somewhat 
younger and therefore has a lower mass-to-light ratio, the mass difference 
may be even greater. The Quadrant arc in the LMC has an estimated mass of 
$3\times10^5$ \msun\ \citep{ee98} and the entire Constellation III may have 
a mass of some $10^6$ \msun, again about an order of magnitude less than the 
NGC~6946 complex.

  From this comparison, it is clear that the NGC~6946 complex presents
a quite unique case, even just by its high concentration of very young
stars and clusters. Other peculiar features include the luminous young
super-star-cluster and the sharp western rim.  Ideally, all these peculiar 
features should be explained jointly, possibly by some unusual process of 
star and cluster formation.  

\subsection{The intrinsic shape of the complex}
  
  The complex has a very sharp and regular outer rim, especially towards the 
west.  The only other known features of such  shape are the Quadrant (with the 
same radius, $\sim300$ pc, and age, $\sim15$ Myr) and Sextant arcs in the LMC 
\citep{ee98}. But what is the true 3-dimensional shape of the complex?  
Figure~\ref{fig:circle_ell} shows NOT images displayed at high contrast to 
make the boundary between the complex and its surroundings more clear. The 
ellipse corresponds to a circular, but planar feature in the inclined 
disk of NGC~6946. 
Overall, the outer boundary of the complex appears to be better fitted by the 
circle than by the ellipse, suggesting a spherical rather than ring-like 
geometry (Efremov 2000; 2001), but note that only the western boundary is 
sharply defined. Towards the east the complex merges with the
spiral arm and south of it are a number of young star forming regions embedded
in H$\alpha$ emission, one of which (at the south-east) has an arc-like shape 
that is unlikely to be an artifact of light absorption, and was sketched 
by Hodge (1967).  The W boundary of the complex by itself 
does not provide strong constraints in favor of either a circular or 
elliptical projected shape. 
  
  Of course, inferences about the 3D shape of the complex based on its 2D 
appearence rely heavily upon the assumption of either spherical or ring-like 
symmetry. It is difficult to imagine how a spherical complex with a diameter 
of about 600 pc could form in a presumably much thinner disk.  A similar
problem exists in the LMC for the Quadrant and Sextant arcs which also have
have circular projected shapes, but larger radii than the thickness of the
LMC gas disk (Efremov 2001a).  The scale 
height of the Milky Way thin stellar disk is only about 100 pc \citep{gil89} 
and the FWHM thickness of the molecular gas disk is $\approx 140$ pc 
\citep[e.g.][]{bro88}.  A thickness of 220 pc has been found for the molecular 
gas disk in the edge-on spiral NGC~891 \citep{sco93}.  The nearly face-on 
orientation of NGC~6946 makes it difficult to obtain information about the 
thickness of its disk, but \citet{eh93} presented a consistent model for 
Faraday rotation, depolarization and thermal radio emission in which most of 
the ionized gas in NGC~6946 resides in clouds within a disk of $\sim100$ pc 
full thickness.  Thus, it appears likely that the disk of NGC~6946 has a 
thickness of 100--200 pc and that much of the material in the complex is 
contained within about this height, implying a somewhat oblate intrinsic shape.

\subsection{Origin}

  Our result that the complex has experienced two major star formation episodes 
clearly depends on the reliability of the stellar isochrones.  However, 
multiple episodes of star formation separated by a few Myr are also known in 
other nearby starbursts such as 30 Dor \citep{rub98,sel99,wal99}, although on 
smaller scales, and might be a result of feedback mechanisms such as supernova 
explosions and winds from massive stars.  While the \citet{trew98} map shows 
less extinction in the Hodge complex compared to the surroundings, presumably 
indicating that the complex forms a cavity cleared out by stellar winds and 
supernovae over the past 25--30 Myr or more, we still see clear indications of 
substantial reddening variations across the complex, ranging from essentially 
no reddening to a $B$-band absorption of 1--2 mag or more. The regions of 
highest absorption are along the outer arc and near the two dark clouds 
identified in Paper I.  In some of the denser parts of these clouds, star 
formation could still be taking place. In the following we briefly discuss a 
few possible scenarios for the formation of the complex, but we refer to 
Paper I and \citet{efr01} for further details.

\subsubsection{Formation by collapse of spiral arm gas}

  In Paper I it was shown that supernova explosions within the complex could 
have provided enough energy to clear out any residual gas and create the
current bubble-like appearance.  However, a more puzzling 
question is how such a large-scale star forming event was initiated in the 
first place at a relatively isolated location within NGC~6946.  The location 
at the end of a spiral arm may be a clue to this question, and in Paper I it 
was suggested that the complex originated in a large-scale asymmetric collapse 
of gas at the end of the spiral arm. If the overall star formation efficiency 
was $\approx10$\% (typical for star formation in galactic Giant Molecular 
Clouds) then the initial gas mass must have been around $10^8$ \msun.  For a 
spherical cloud with a radius of 300 pc this corresponds to a mean density of 
$n(H_2) \approx 20 \, {\rm cm}^{-3}$, or higher for a flattened geometry.  
Thus, the proto-cloud may have resembled the ``Super Giant Molecular Clouds'' 
proposed by \citet{hp94} as the progenitors of globular clusters.  Fig.~1 in 
Paper I shows regions of star formation at similar positions with 
respect to other spiral arms in NGC~6946, although none of them have quite as 
striking an appearance as the complex discussed here.  However, most of these 
other regions are still dominated by H$\alpha$ emission, indicating younger 
ages than for the Hodge complex. The peculiar appearance of the latter may be 
related to the clearing of gas, providing a better view of its stellar contents.

  The high density of the YSSC is not surprising considering the environment
in which it was born.  The star formation history in this region suggests that 
15 Myr ago there was a bright association of stars with a total mass of around 
$4\times10^6$ M$_\odot$, from the integral under the curve of the oldest burst 
in Figure \ref{fig:sfh}.  This mass corresponds to an enormous luminosity and 
energy output in the form of HII regions, stellar winds and supernovae, but it 
is not unusual compared to other giant star complexes in the outer spiral arms 
of galaxies.  For example, the supergiant HII regions in the study by 
\citet{ken84} contain over 1000 OB stars; considering the IMF, this corresponds 
to over $10^6$ M$_\odot$ of total stars.  The stellar energy from the 
association in NGC 6946 would have immediately begun to clear a cavity around 
it, eventually leading to the big bubble seen today.  At the same time, it 
would have compressed any residual molecular material and triggered more star 
formation.  Because the formation of an OB association is relatively 
inefficient, there was probably a few times $10^7$ M$_\odot$ of dense gas left 
over within 100 pc or so of the first generation stars.  If about $10^7$ 
M$_{\odot}$ of this residual material was in the form of one large cloud, then 
the compressed part of this cloud facing the association core would have made 
one large and dense cluster, which is now the YSSC. A similar phenomenon, 
scaled down by a factor of $\sim100$, occurred in the Orion association, with 
the Trapezium cluster triggered in the comet-shaped head of a GMC \citep{bal87} 
compressed by the older Orion association.  The triggered stars in NGC 6946 
formed with an equally high efficiency as the Trapezium cluster, leaving a 
bound remnant, and then broke apart its remaining GMC into the two pieces that 
are now seen as two dark clouds, one to the North and the other to the East 
of the YSSC. A faint arc of extinction connects these two clouds giving the 
impression of a partial shell centered on the YSSC.  At this time, the combined 
pressures from the aging first generation of stars, which is getting more and 
more dispersed with time, and the YSSC drove the expansion of the bubble 
further and triggered more star formation. This third generation of stars had 
a more distributed pattern than the YSSC, forming a lot of smaller clusters,
because the clouds were more dispersed when they formed.  It is evident as the 
young burst in figure \ref{fig:sfh}. One feature which remains unexplained in 
this scenario is the sharp western rim; however, it is not clear to what 
extent this is real or an effect of dust obscuration.

\subsubsection{High velocity cloud impact}

  Another mechanism which might explain the formation of the complex is 
the infall of a massive high velocity cloud.  Some support for this idea
is provided by the fact that NGC~6946 is known to have a large number
of such clouds projected on its disk \citep{bv92,kam93}.  The velocity of 
the YSSC and of the HII gas inside and near the complex being on average 
the same and quite similar to the rotational velocity of the galaxy at 
this position \citep{efr01}, one has to conclude that the infall was oblique, 
i.e.\ the trajectory of the infalling cloud had a small angle to the galaxy 
plane.  Note also that the Eastern edge of the complex is neither sharp 
nor circular  and the overall shape of the complex resembles a comet with 
a very short Eastward tail (see Figure 7 in Paper I).  The shape of the 
complex in comparison with simulations of partial spheres seen under 
different angles indicates an infall angle of about 10--20 degrees or so 
\citep{efr01a,efr01b}. This suggestion seems to be compatible with the 
comet-like shape of the complex, which may indicate a Westward movement of 
the impacting cloud. The radial velocity of the YSSC and the
bulk of the surrounding ionized gas is about 150 km/s \citep{lar01,efr01},
whereas the average rotational velocity of surrounding HII gas is about
130 km/sec. This is compatible with the infall hypothesis.  We note that 
the infall of a high-velocity cloud has also been suggested as the 
formation mechanism for the Gould belt \citep{com94}.

\subsubsection{Super-explosions}

  Finally, we consider the possibility originally suggested by \citet{hod67}, 
namely a highly energetic explosion, which might have triggered star
formation in shells of swept-up gas and generated the arc-like structures.
Some support for this idea is found in data on the kinematics of HII gas in 
the complex, which display two young expanding HII shells with different 
expansion velocities and dynamical ages of a few Myr \citep{efr01}. Formation 
of some of the younger stars 
might have been triggered by hypernova progenitors ejected from the YSSC 
\citep{efr01a}, although the formation of the older stellar population in 
the complex and the YSSC itself is not easily explained as due to such 
super-explosions.

\subsubsection{The young super star cluster}

  Noting the high density of point sources near the YSSC, it was suggested
in Paper I that the YSSC might have formed from accretion of a number of 
less luminous clusters surrounding it. Fig.~\ref{fig:clmap} shows 4 clusters
(\#865, \#975, \#1094 and \#1236) near the YSSC, but it is now clear that
many of the luminous objects near the YSSC are probably not clusters, but 
individual $\la6$ Myr old stars. Thus, the case for formation of the YSSC
by coalescence of smaller clusters may have been weakened.  The cluster may 
simply have formed by normal star formation mechanisms, albeit on a much 
larger scale than is seen in the Milky Way. 

  In addition to the $A_B=1.48$ foreground absorption, the reddening map 
indicates an internal absorption of $A_B=0.53$ at the position of the
YSSC, or $A_B$(total) = 2.01 mag. Alternatively, the $Q$-method \citep{vh68} 
gives an internal absorption of $A_B = 1.04$, or $A_B$(total) = 2.52.
The cluster might therefore be as bright as $M_V\approx-14$. 

  Although the YSSC has an unusually high luminosity (and mass) for a young 
star cluster in a normal spiral galaxy, other similarly luminous clusters
are known elsewhere.  Similar bright clusters exist in galaxies such as M51,
M83 and NGC2997 \citep{lr99,lar00}, some of which are nearly as bright as the 
YSSC in NGC~6946, and the 2nd brightest cluster within NGC~6946 is only about
1 magnitude fainter than the YSSC.  The cluster is also quite comparable to 
the most luminous clusters in starburst galaxies \citep[e.g.][]{whit01}.
Characteristic for all these bright clusters is that they are found in 
environments of high star formation activity. 

  It is interesting to consider how the complex fits into the SFR vs.\ 
specific cluster luminosity relation discussed by \citet{lr00}. From the
photometry in Table~\ref{tab:cltab} and Paper II we obtain the specific 
$U$-band luminosity $T_L(U)$ for the YSSC with respect to the complex as
\begin{equation}
  T_L(U) = 100 \, L_{\rm YSSC} / L_{\rm complex} = 18
\end{equation}
This is among the highest $T_L(U)$ values seen anywhere. Again, internal
reddening has not been taken into account, but since this should affect both
the cluster and the complex we have simply chosen to use the directly
observed luminosities.
Actually, since the colors of the YSSC and the complex as a whole are
very similar, it doesn't matter much in which band the specific luminosity
is computed in this particular case.  For a total complex mass of $10^7$ \msun, 
radius 0.3 kpc and age 15 Myr, the star formation rate surface density 
\ssfr\ = 2.4 \msun\ yr$^{-1}$ kpc$^{-2}$, even higher than any galaxy studied
by \citet{lr00}.  It should be noted that they defined $T_L(U)$ for a 
cluster system with respect to the integrated luminosity of the entire 
underlying galaxy, which includes some contribution from older stellar 
populations even in $U$. It is also not obvious that the same \ssfr -- $T_L(U)$ 
relation observed for galaxies is valid locally. 

Nevertheless, the problem 
of understanding the presence of the YSSC in the NGC~6946 complex is perhaps 
no different from understanding the presence of massive star clusters in 
actively star forming environments in general. With an initial gas supply
of some $10^8$ \msun\ within a radius of 300 pc, collecting $\sim10^6$
\msun\ in a dense core and forming a massive cluster there does not seem
difficult.  The fact that such massive clusters are not observed in the 
Gould belt might simply be due to the much lower star formation density there.

\section{Summary and Conclusions}

  We have presented HST WFPC2 data for a peculiar stellar complex in NGC~6946.
The complex is dominated by a young super star cluster which accounts for
about 18\% of the total luminosity of the complex. The linear dimensions 
are similar to those of the local Gould belt, but the total stellar mass 
contained within the NGC~6946 complex is 1 -- 2 orders of magnitude higher. 
The complex thus has the characteristics of a (very localized) starburst, 
which might explain the presence of a very luminous cluster within it. The
cluster itself presumably formed in the same way as massive clusters elsewhere,
in the dense core of a large concentration of molecular gas.  In addition 
the the YSSC, the complex also contains a number of clusters with more 
modest luminosities; however, about a dozen of these are still comparable to 
the brightest young clusters in the Milky Way.

  We have demonstrated that it is possible to model the CMD of stars in the 
complex by a superposition of currently available stellar isochrones.  The 
best match is obtained for solar-metallicity models from the Padua group.
From the CMD modeling we have found tentative evidence for two star forming 
episodes within the complex. The time of onset for the first burst is not well 
constrained in our data but may have occurred some 25--30 Myr or longer ago.  
This burst lasted until 15--20 Myr ago and was followed first by the formation
of the YSSC, and then by a quiescent period.
The second burst began about 6 Myr ago and may be continuing until the present 
day.  However, the apparent age gap relies on the model prediction 
that a 10--20 Myr old population should produce a substantial number of 
red supergiants brighter than $M_V=-5$, which are not observed.  If no RSGs 
are produced in this age range, or if they are fainter than predicted, then 
an age gap might not be required.  Stars 
in the corresponding mass range, 12 -- 18 M$_{\odot}$, are not usually 
believed to become Wolf-Rayet stars \citep{mc94}, although it has been 
suggested that stars in M31 with masses as low as 13 -- 15 M$_{\odot}$ might 
become W-R stars, spending at most only a brief period as RSGs \citep{mas98}.  
Clearly, a more detailed analysis has to await a better understanding of 
the evolution of massive stars, but in the meantime the NGC~6946 complex might
serve as another testbed for such studies. Some support for two episodes of
star formation is also seen in the age distribution of stellar clusters.

  The circumstances leading to the formation of the complex remain elusive,
but we have briefly discussed a few possible scenarios: 1) Collapse of a 
large concentration of spiral arm gas, as suggested by \citet{eel00}. This 
initially formed a $4\times10^6$ M$_{\odot}$ OB association, which 
subsequently compressed
any residual gas and triggered further star formation, including the
YSSC;  2) impact by a high-velocity cloud, suggested as the formation 
mechanism for the Gould belt \citep{com94}; 3) Shock compression of a gas 
cloud by an energetic super-explosion, following the suggestion by 
\citet{hod67}.

  The best way to make further progess in the understanding of this peculiar
stellar complex would be high resolution HI and CO data. The gas
distributions and velocities might still have imprints of the event
which formed  the complex. Existing data for the gas show no peculiarities,
except for two expanding young HII supershells (Efremov et al.\ 2001), which 
are probably connected with the younger stellar generation. Imaging at 
near-infrared wavelengths would also help in penetrating through regions
of high extinction, providing a clearer view of the true morphology of 
the complex.
  
\acknowledgments

  Support for Program number GO-8715 was provided by NASA through
grants GO-08715.02-A and GO-08715.05-A from the Space Telescope Science 
Institute, which is operated by the Association of Universities for 
Research in Astronomy, Incorporated, under NASA contract NAS5-26555. 
SSL acknowledges support by National Science Foundation grant 
number AST9900732. Yu.E. is grateful for support by grants from
the Russian foundations, 00-02-17804 and 00-15-96627. EJA  acknowledges 
the partial support from DGICYT through grant PB97-1438-C02-0 2 and by 
the Research and Education Council of the Autonomous Government of 
Andalusia  (Spain).  Finally, we thank the referee for insightful comments.

\clearpage

\begin{figure}
\figcaption{\label{fig:n6946rgb}
  A color composite image showing the complex centered on the PC chip.
The image is an approximation 
to a true-color image and was assembled from F336W and F439W (blue), 
F555W (green) and F555W, F814W and F656N (red).  HII regions appear red.
}
\end{figure}

\begin{figure}
\figcaption{\label{fig:clmap}
  An F555W ($V$-band) image of the stellar complex showing the location 
of cluster candidates, i.e.\ objects with $M_V<-7.5$ and FWHM$>\szcut$ pixels.
The arrow indicates the North and East directions. 
}
\end{figure}
\clearpage

\begin{figure}
\plotone{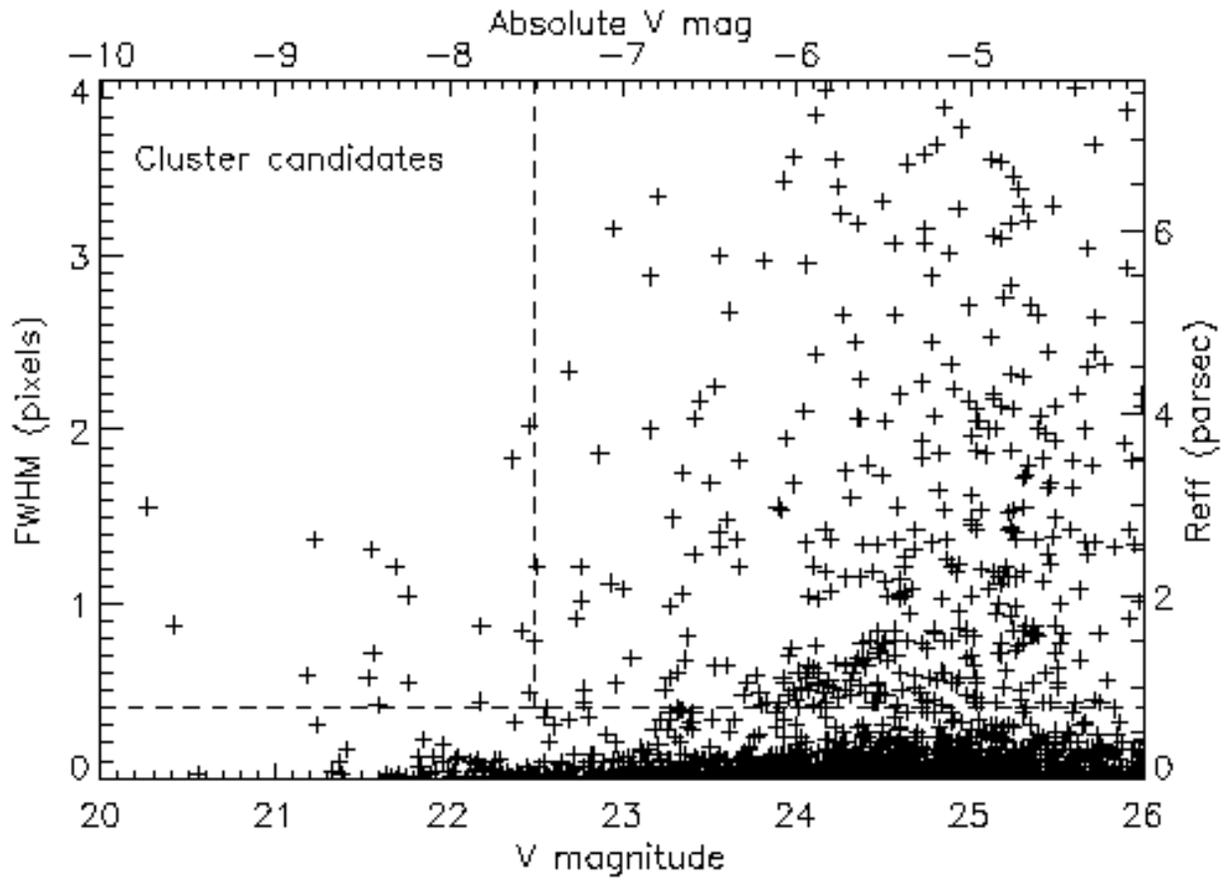}
\figcaption[f3.eps]{\label{fig:V_sz}
  Object size as a function of $V$ magnitude for all objects in the PC
  frame. Stellar objects are selected as objects with FWHM $< \szcut$ pixels.
}
\end{figure}
\clearpage

\begin{figure}
\plotone{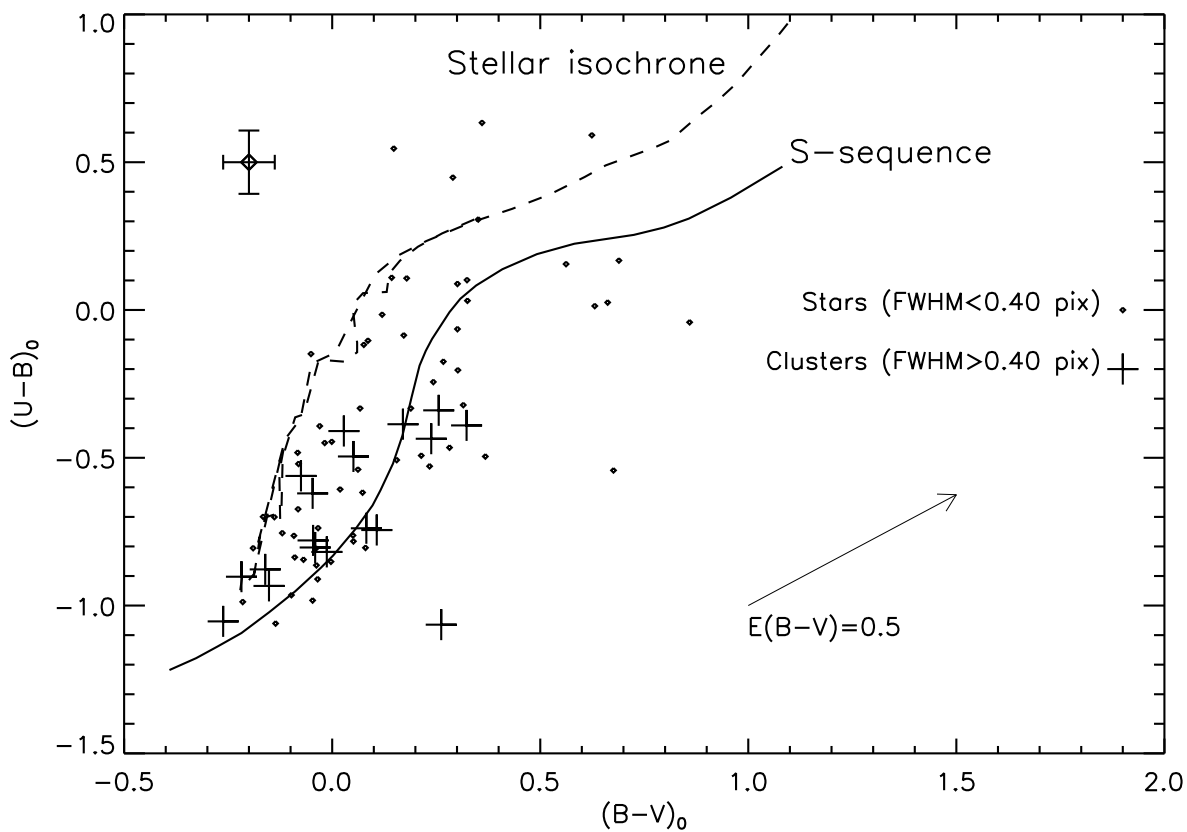}
\figcaption[f4.eps]{\label{fig:bv_ub}
  (\bvz, \ubz) two-color diagram for objects brighter than $V=22.5$ 
  ($M_V=-7.5$) (aperture photometry), corrected for foreground reddening.  
  Point sources (FWHM$<$\szcut\ pixels) are 
  shown with diamond symbols. The solid line is the Girardi et al.\ (1995) 
  ``S''-sequence, while two isochrones from Girardi et al.\ (2000) 
  (4 and 10.0 Myr) are shown with dashed lines.  The arrow indicates the
  reddening vector corresponding to $E(B-V) = 0.5$ mag.
}
\end{figure}
\clearpage

\begin{figure}
\plotone{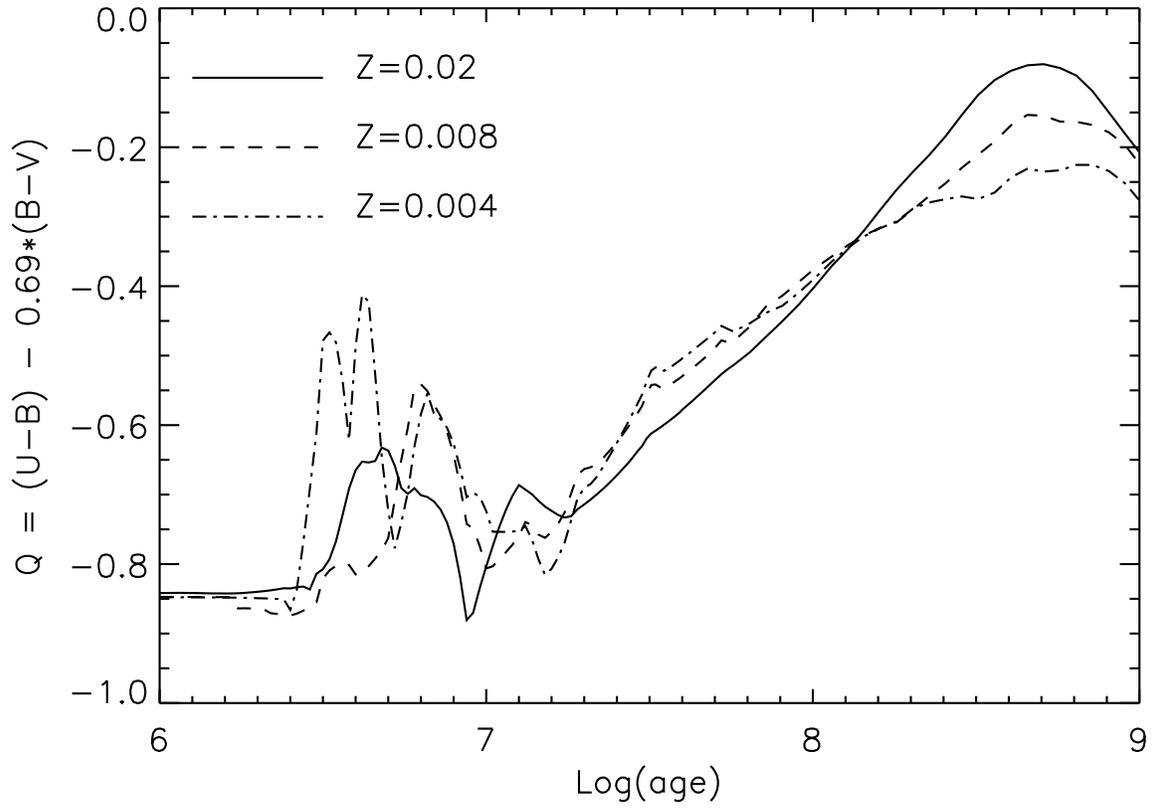}
\figcaption[f5.eps]{\label{fig:qplot}
  Evolution of the reddening-free $Q_1$ index as a function of age for
$Z=0.02$ (solar), $Z=0.008$ and $Z=0.004$.
}
\end{figure}
\clearpage

\begin{figure}
\figcaption{\label{fig:abmap}Reddening map with red and blue
indicating high and low reddening, respectively. The wedge at the bottom
indicates a range from $A_B=1.5$ to $A_B=3.5$.  Individual reddenings
were determined from the $UBV$ colors of stars and the map was produced by 
smoothing with an Epanechnikov kernel.}
\end{figure}
\clearpage

\begin{figure}
\plotone{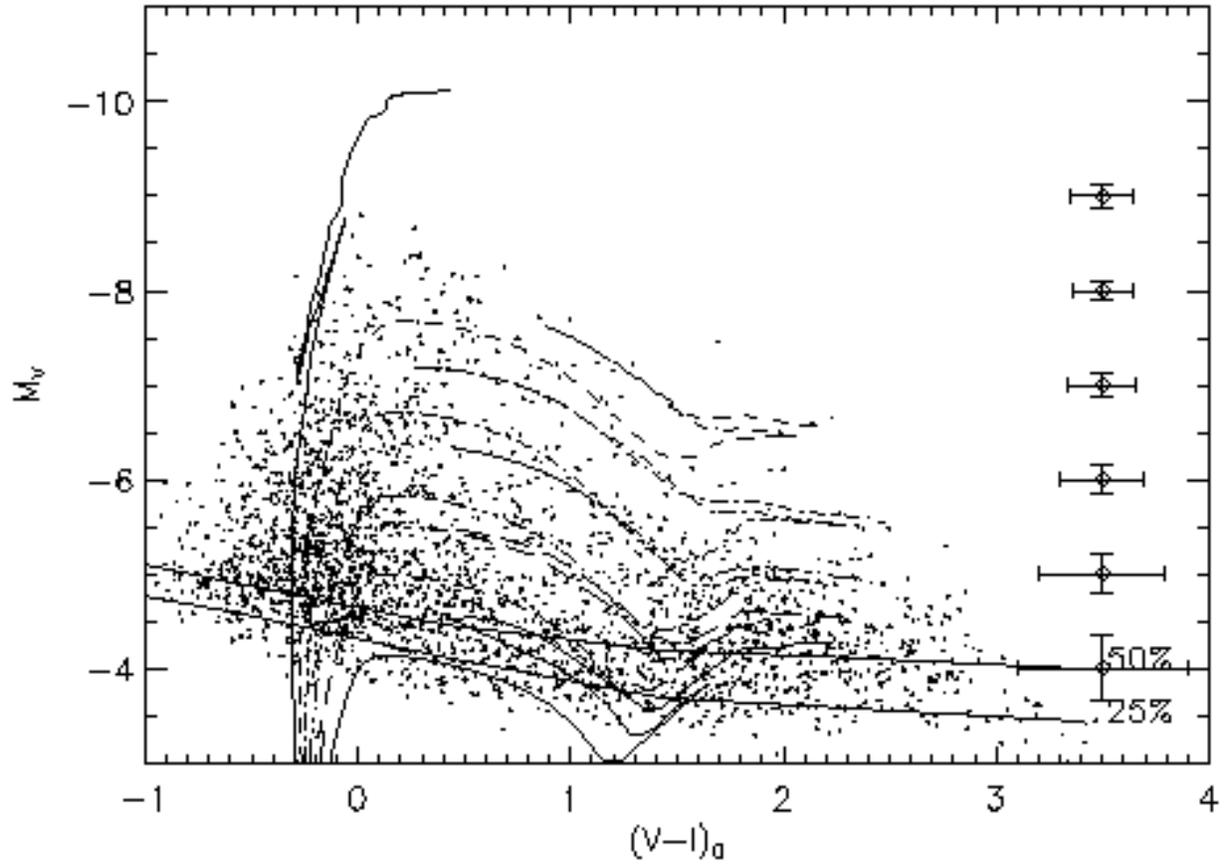}
\figcaption[f7.eps]{\label{fig:cmd_iso}
  ($\viz, M_V$) color-magnitude diagram for PSF-fitting photometry for point 
  sources in the PC frame. Stellar isochrones for ages of 4, 10, 16, 25, 40 
  and 63 Myr from Girardi et al.\ (2000) are superposed on the plot. The
  50\% and 25\% completeness limits are also shown.
}
\end{figure}
\clearpage

\begin{figure}
\plotone{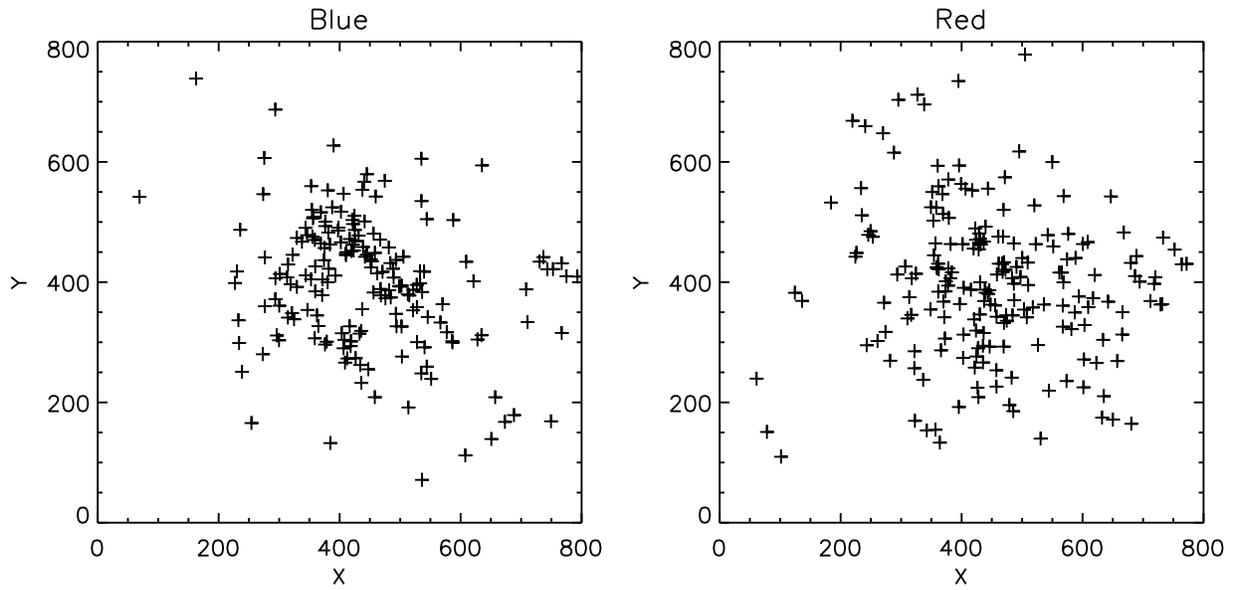}
\figcaption[f8.eps]{\label{fig:spat}Spatial distributions of
early-type stars stars brighter than $V=23.5$ ($M_V<-6.5$) (left) and
red supergiants (right). The early-type (main sequence and supergiant) 
stars appear to be more concentrated towards the center of the complex.
}
\end{figure}
\clearpage

\begin{figure}
\plottwo{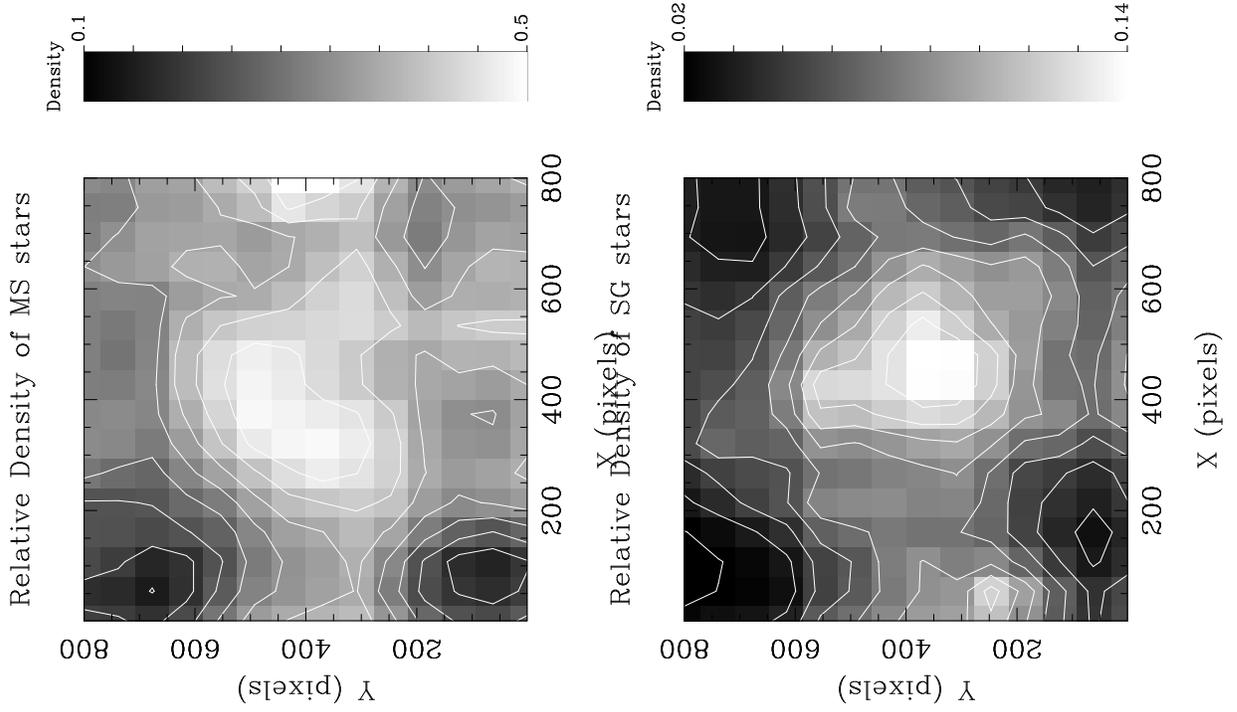}{f9b.eps}
\figcaption[f9a.eps,f9b.eps]{\label{fig:densipop}Relative density 
distributions for early-type and red supergiant (RSG) stars.  As in
Fig.~\ref{fig:spat}, the maximum in the relative number density of red stars 
is offset from that of the blue stars and coincides with the center of 
curvature of the western rim.
}
\end{figure}
\clearpage

\begin{figure}
\plotone{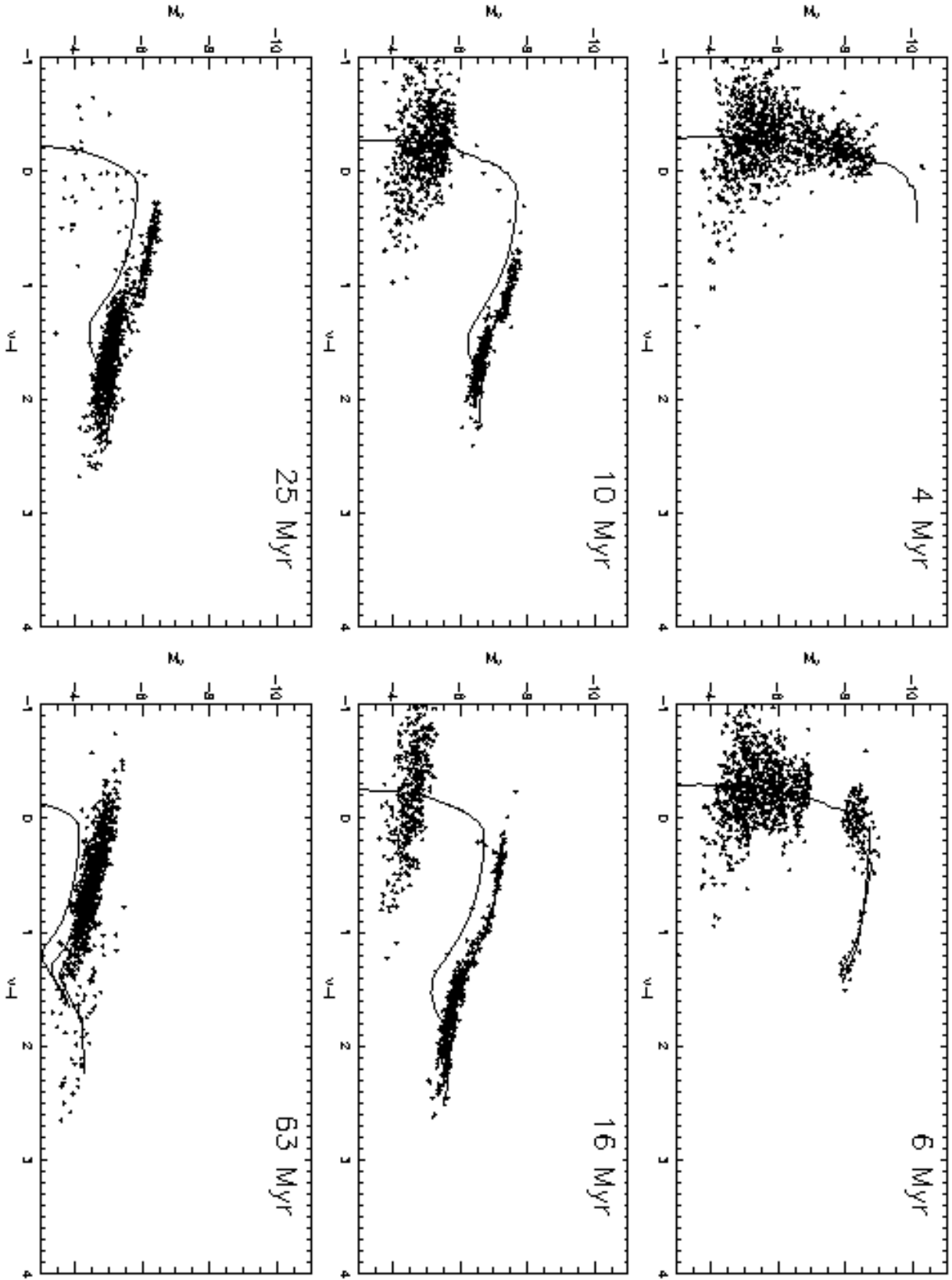}
\figcaption[f10.eps]{\label{fig:isofig}Simulated color-magnitude diagrams
  for stellar populations of six different ages, based on Padua isochrones. 
  The isochrones have
  been populated according to a Salpeter IMF.  Using the data
  corresponding to our NGC~6946 photometry, photometric errors have been
  added and the diagrams have been depopulated according to the
  completeness functions. Each plot contains 1000 points.
}
\end{figure}
\clearpage

\begin{figure}
\plotone{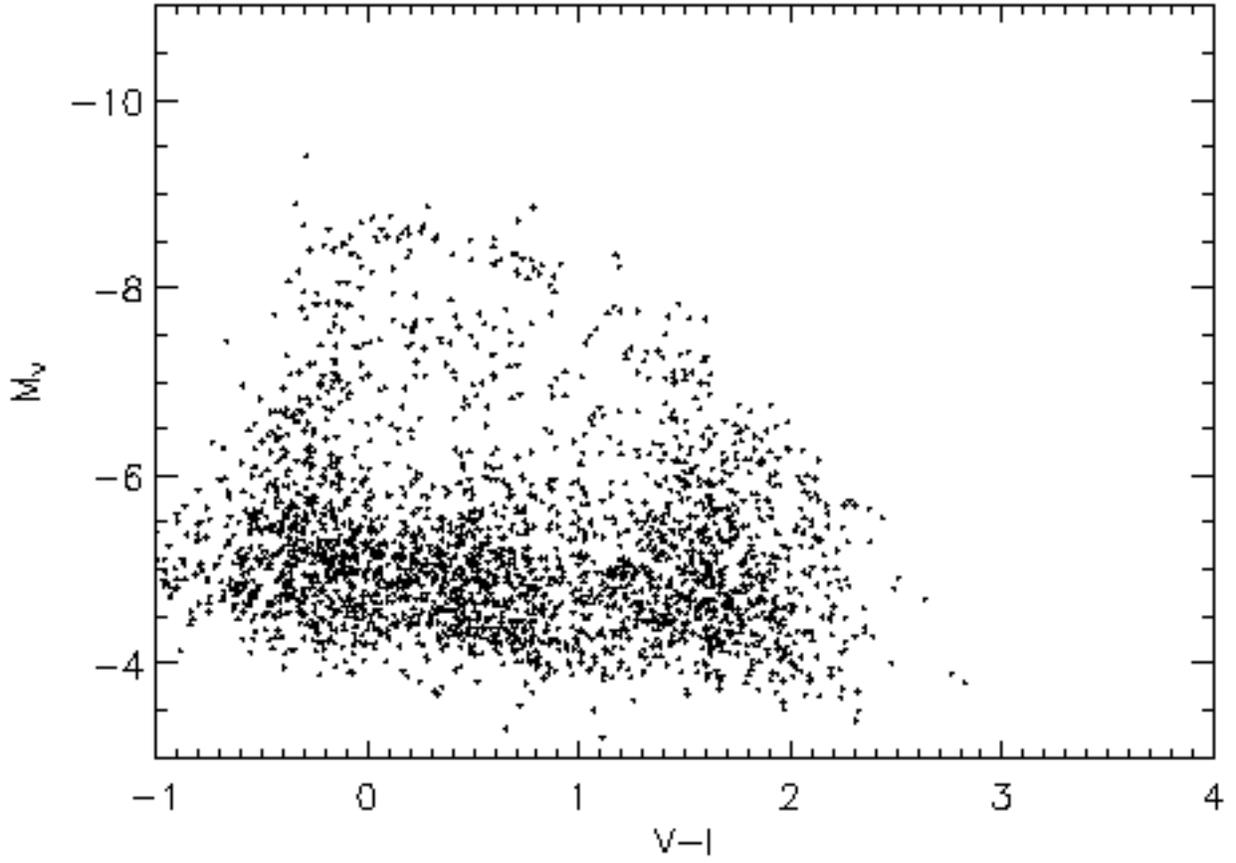}
\figcaption[f11.eps]{\label{fig:synt_cnst_p}Simulated color-magnitude
diagram for constant star formation rate in the range 4.0 -- 100 Myr,
using Padua $Z=0.019$ isochrones.
}
\end{figure}
\clearpage

\begin{figure}
\plotone{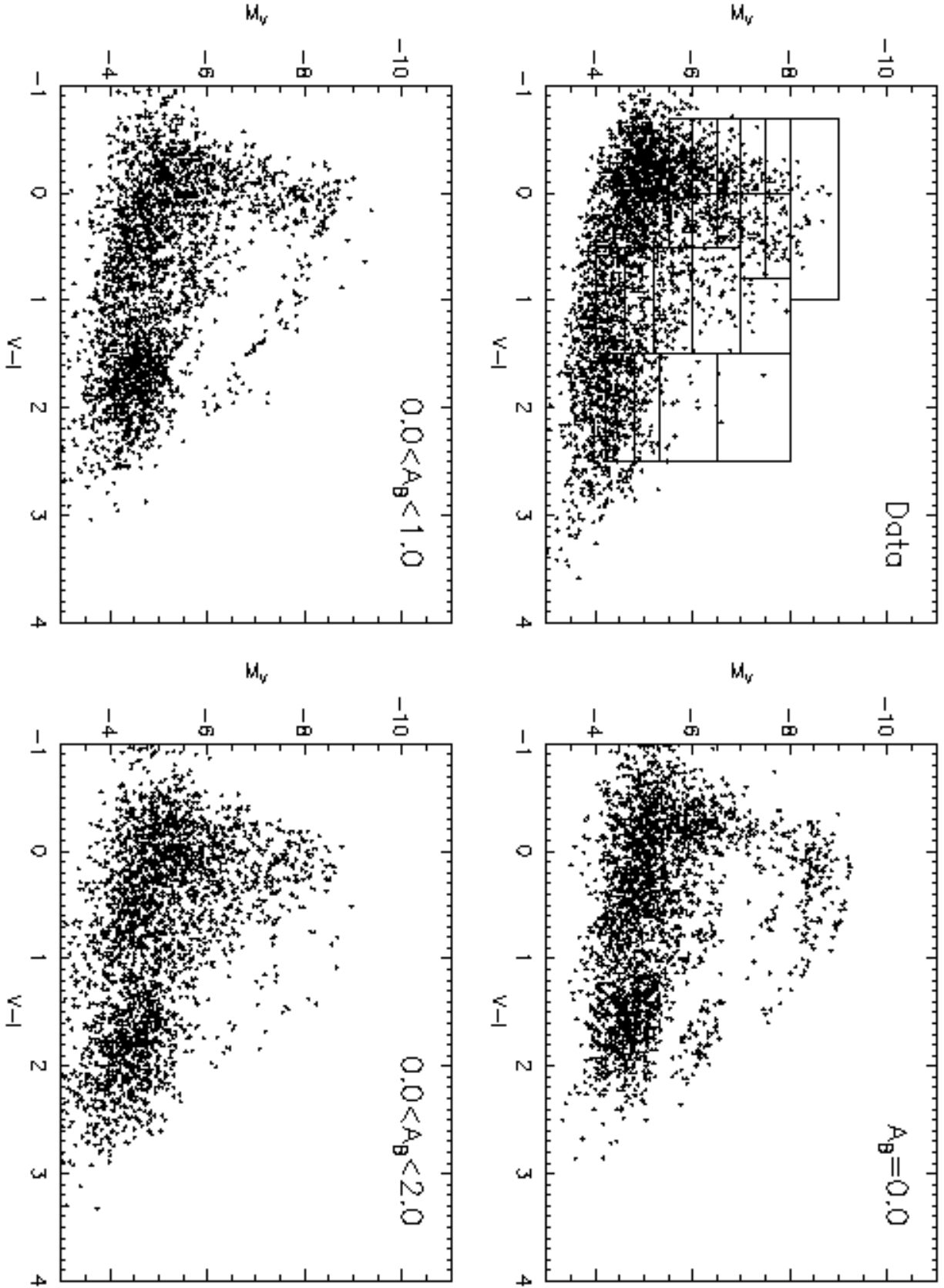}
\figcaption[f12.eps]{\label{fig:syntfig}Simulated color-magnitude diagrams
  for three different reddening distributions. The stars were assigned
  random reddenings in the given intervals.
}
\end{figure}
\clearpage

\begin{figure}
\plotone{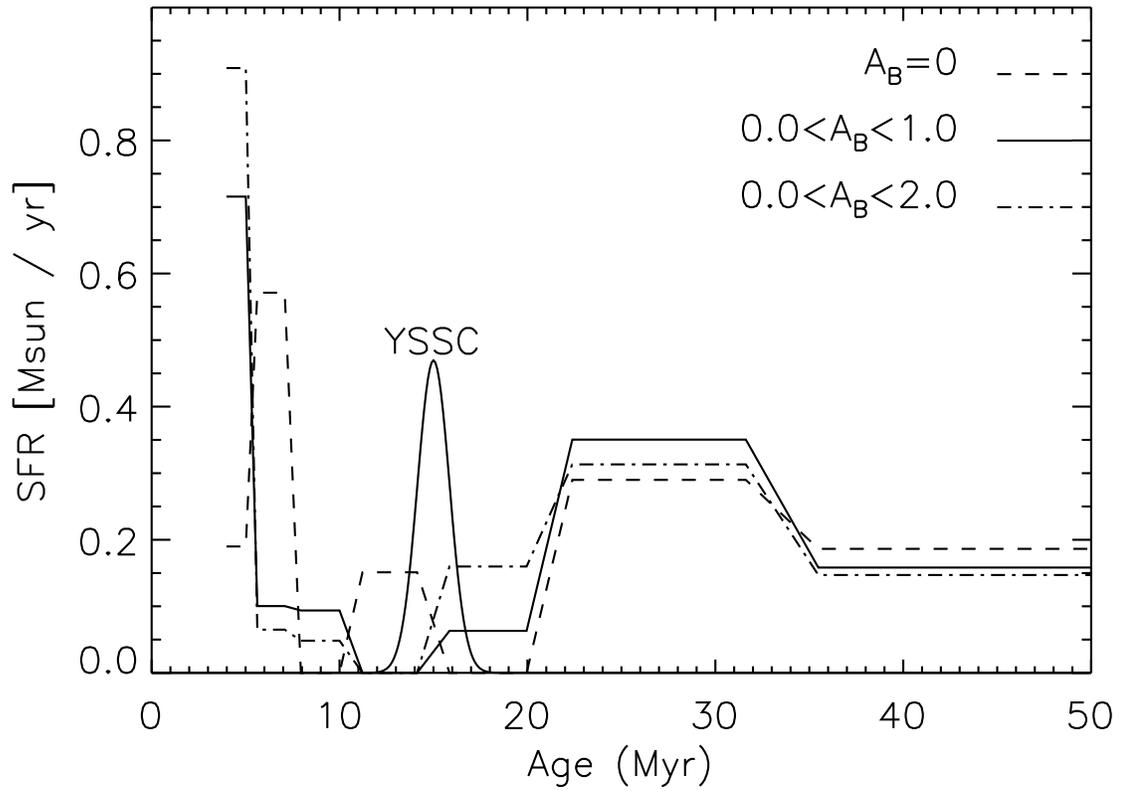}
\figcaption[f13.eps]{\label{fig:sfh}Star formation history derived
from CMD fitting for three different assumptions about internal
extinction in the complex. The SFR has been normalized assuming a total
complex mass of $10^7$ \msun . The Gaussian curve represents the
young super star cluster.
}
\end{figure}
\clearpage

\begin{figure}
\plotone{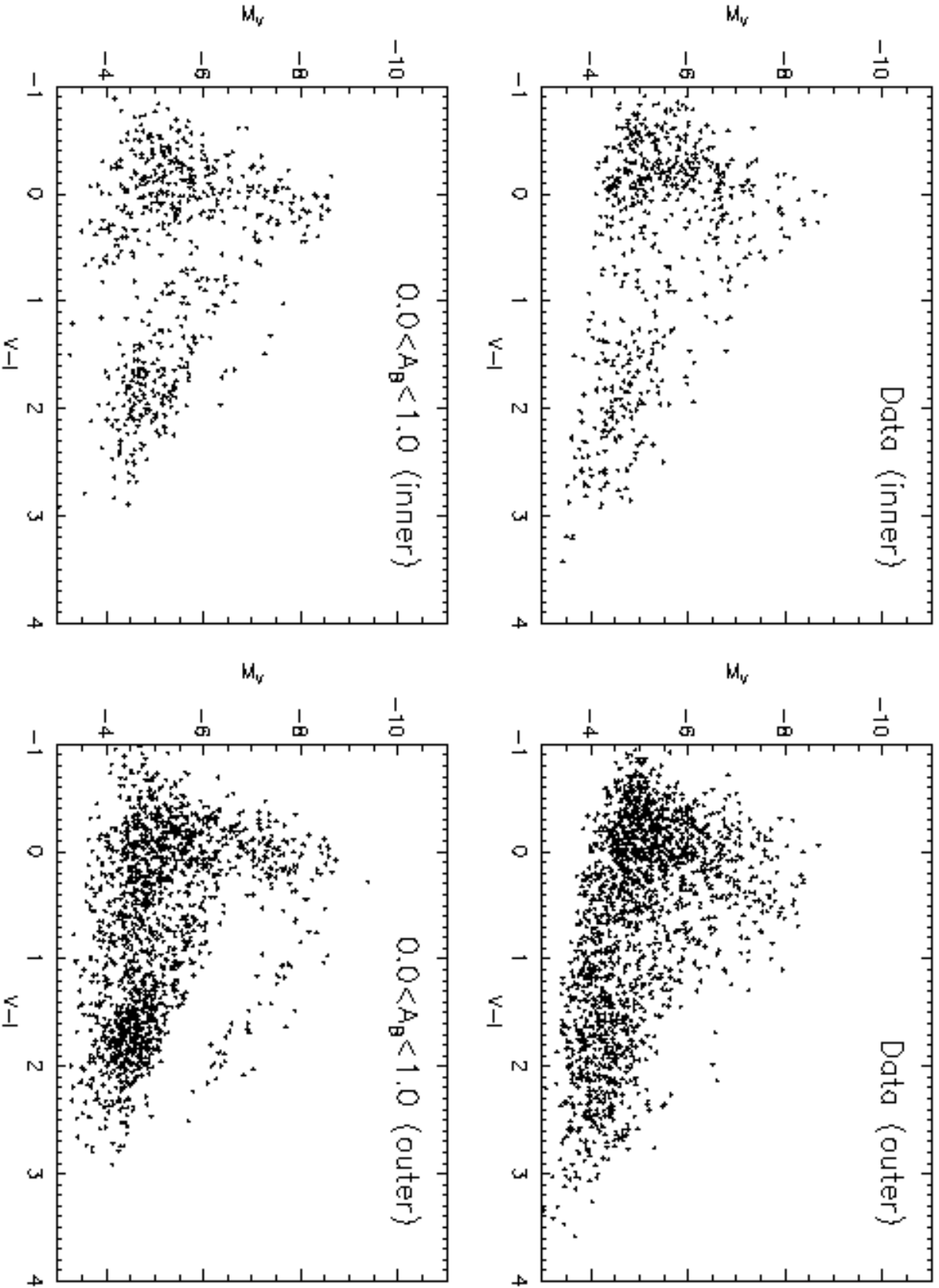}
\figcaption[f14.eps]{\label{fig:syntfig_io}Observed and
  simulated color-magnitude diagrams for the central 100 pixels (left) and 
  for $100<r<300$ pixels (right).
}
\end{figure}
\clearpage

\begin{figure}
\plotone{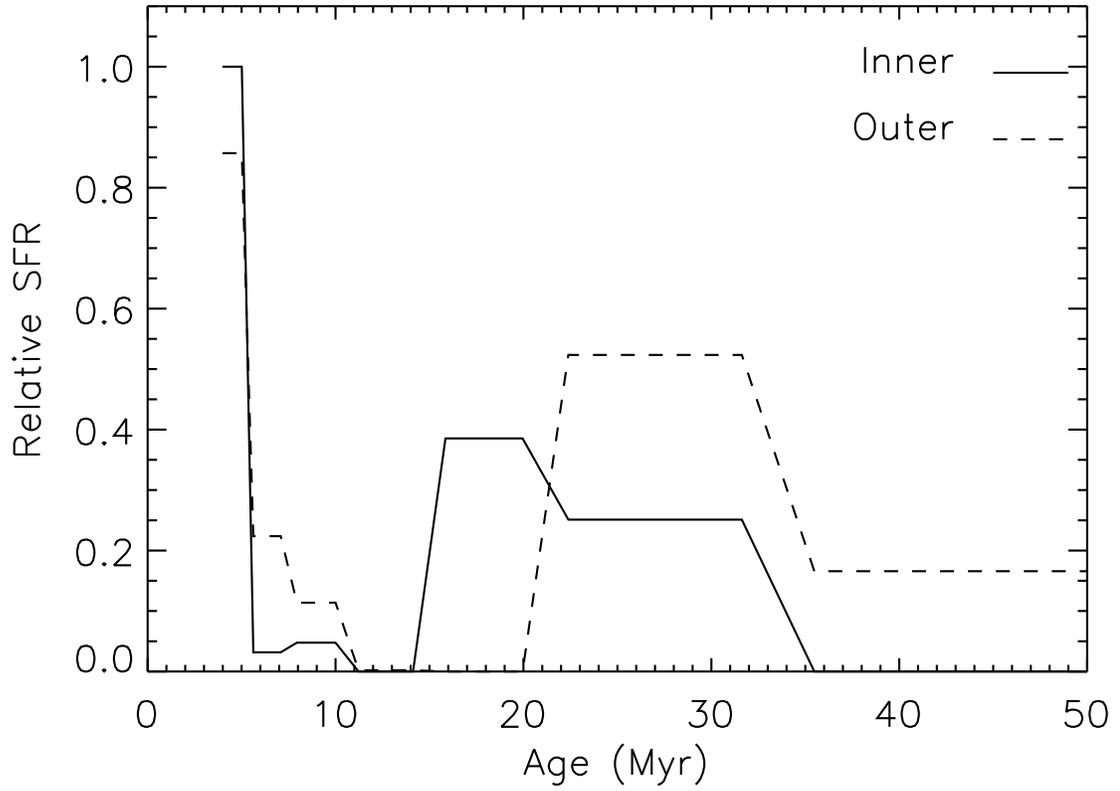}
\figcaption[f15.eps]{\label{fig:sfh_io}Star formation histories for the
central 100 pixels and for $100<r<300$ pixels.
}
\end{figure}
\clearpage

\begin{figure}
\plotone{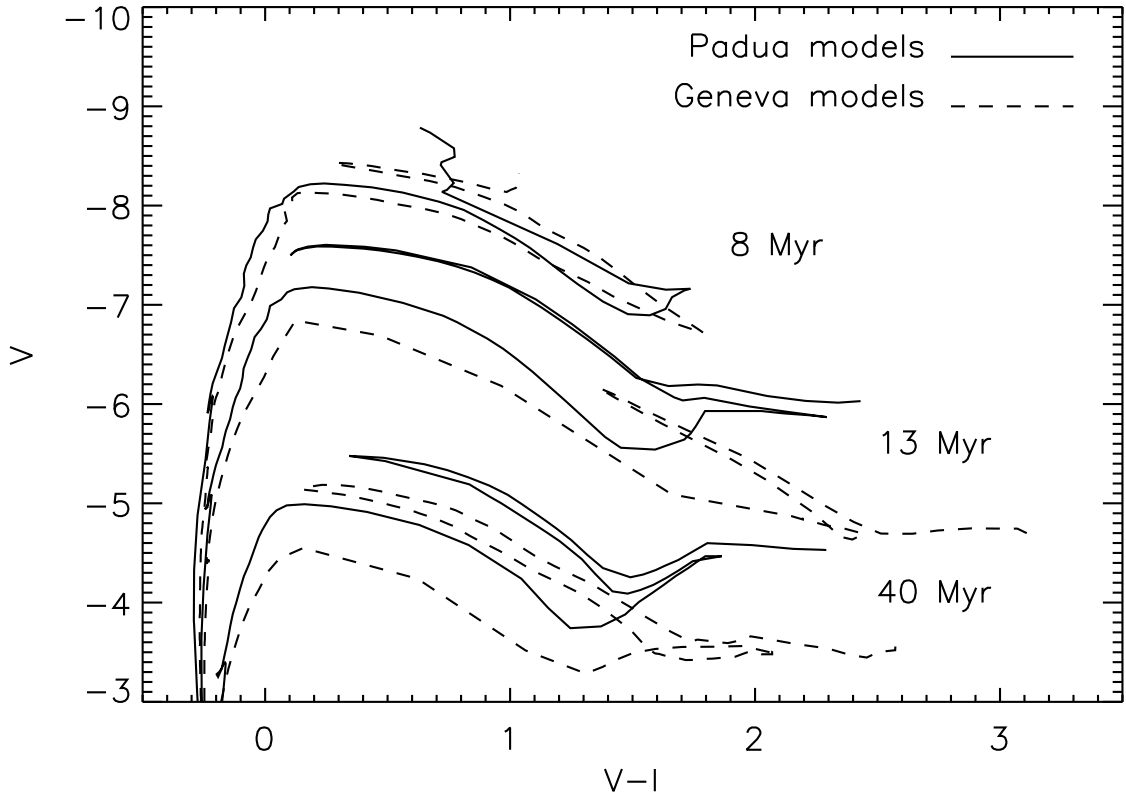}
\figcaption[f16.eps]{\label{fig:isocmp}Comparison of solar-metallicity
isochrones from the Padua group \citep{gir00} and Geneva group \citep{ls01}.
The Geneva isochrones produce cooler and fainter red supergiants for a given
age.
}
\end{figure}
\clearpage

\begin{figure}
\plotone{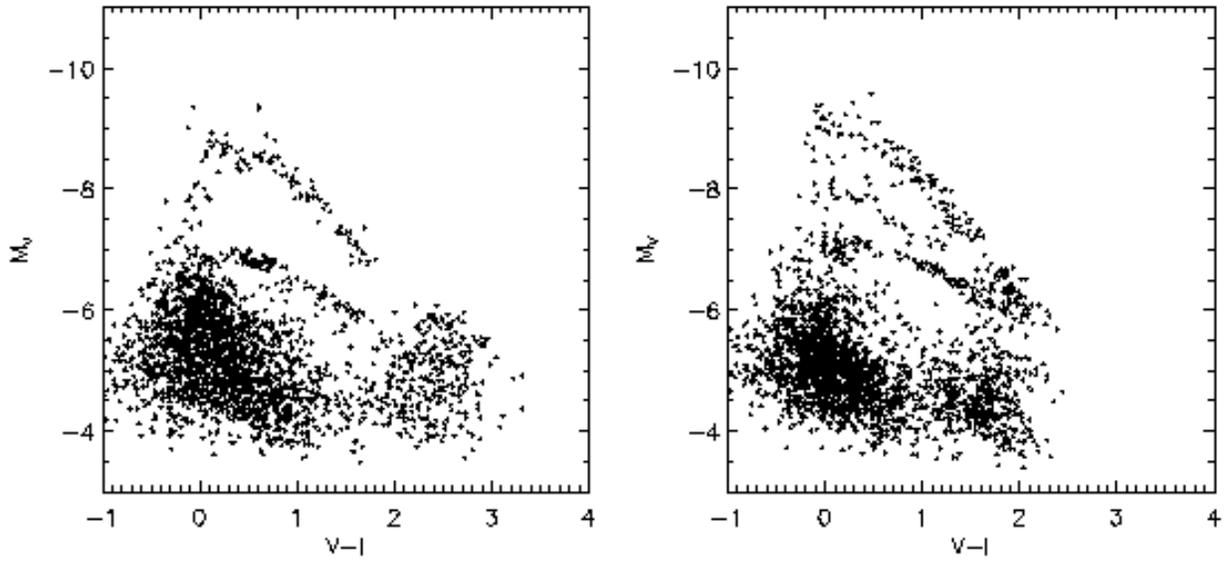}
\figcaption[f17.eps]{\label{fig:synt_cnst_g}Simulated color-magnitude
diagram for constant star formation rate in the range 4.0 -- 100 Myr,
using Geneva $Z=0.019$ (left) and $Z=0.008$ (right) isochrones.
}
\end{figure}
\clearpage

\begin{figure}
\plotone{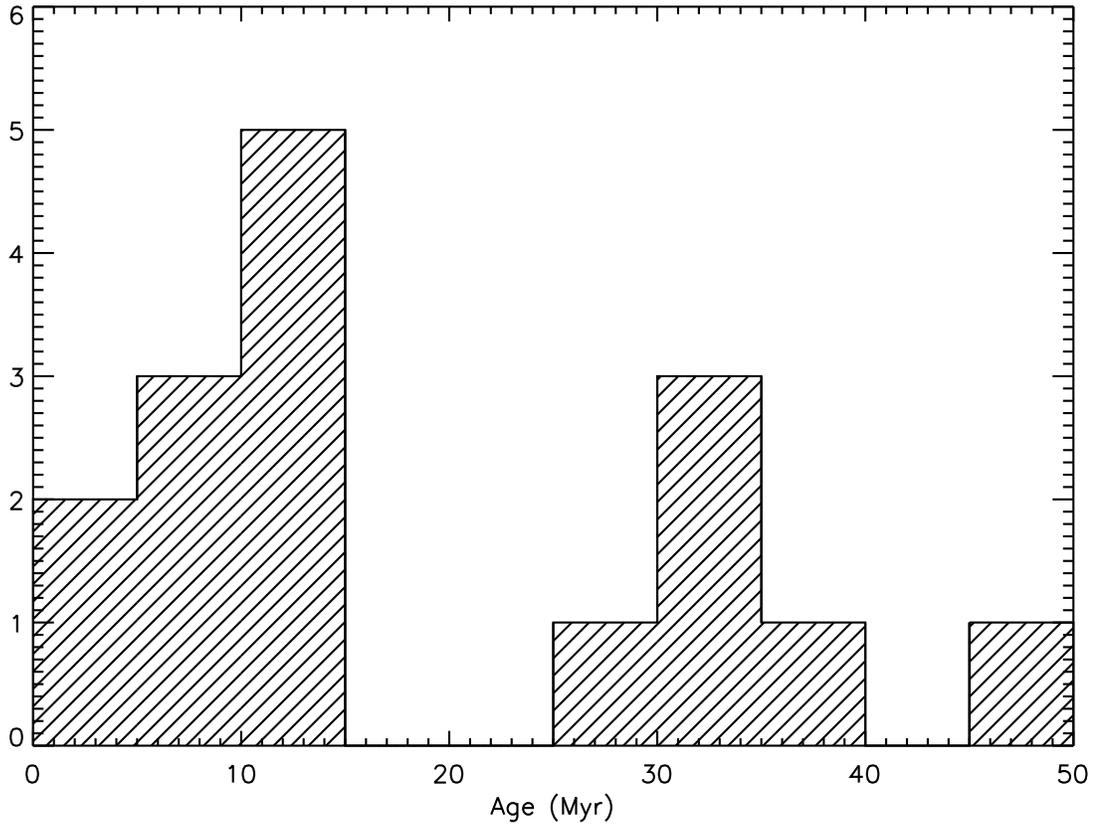}
\figcaption[f18.eps]{\label{fig:clages}Histogram of cluster ages,
determined from de-reddened $UBV$ colors and the Girardi et al.\ (1995)
S-sequence calibration.}
\end{figure}
\clearpage

\begin{figure}
\plottwo{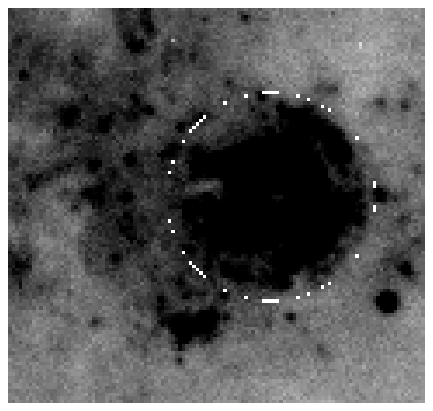}{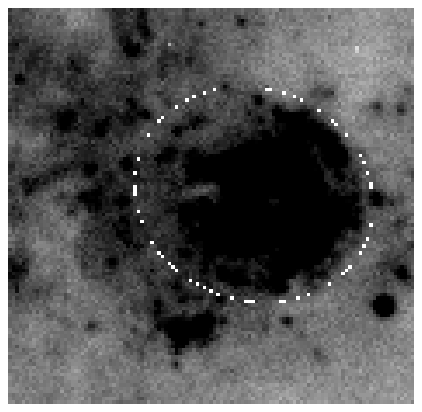}
\figcaption[f19a.eps,f19b.eps]{\label{fig:circle_ell}
  A NOT image of the complex, with circles and ellipses overlaid. North
  is up and east to the left in this figure.
}
\end{figure}
\clearpage

\begin{deluxetable}{lrrrr}
\tablecaption{\label{tab:apc}Aperture Corrections, ($11\rightarrow3$ pixels)}
\tablecomments{Aperture corrections (in magnitudes) are for the PC chip for 
objects with intrinsic FWHM values as listed. Note that the F336W, F439W
and F555W aperture corrections are very similar, independently of object
size.
}
\tablehead{
  Filter  &  \multicolumn{4}{c}{FWHM (pixels / pc)} \\
          &   0 / 0   &  0.25 / 0.32  &  0.50 / 0.64  & 1.00 / 1.29
}
\startdata
 F336W    & $-0.158$   &  $-0.295$    &  $-0.534$   &  $-0.849$ \\
 F439W    & $-0.159$   &  $-0.296$    &  $-0.536$   &  $-0.853$ \\
 F555W    & $-0.152$   &  $-0.304$    &  $-0.547$   &  $-0.867$ \\
 F814W    & $-0.210$   &  $-0.370$    &  $-0.611$   &  $-0.923$ \\
\enddata
\end{deluxetable}

\begin{deluxetable}{lrrrrrc}
\tablewidth{0pt}
\tabletypesize{\footnotesize}
\tablecaption{\label{tab:cltab}
Data for the brightest clusters 
}
\tablecomments{$M_V$ values are measured in $r=11$ pixels (0\farcs5)
aperture. Colors in $r=3$ aperture, with aperture corrections to $r=11$
pixels. Log(age) are from the S-sequence calibration. In addition to
the intrinsic scatter of 0.14, typical uncertainties on log(age) are
about 0.3.}
\tablehead{
  Object & $x,y$ & $M_V$ & \ubz & \bvz & \viz & $Q_1$ 
}
\startdata
Complex & 400,400 & $-15.0$ & $-0.76$ & $-0.05$ & 0.36 & $-0.72$ \\
YSSC    & 520,380 & $-13.2$ & $-0.72$ &   0.10  & 0.59 & $-0.80$ \\
99 & 103,112 & $-7.99\pm0.06$ & $-0.41\pm0.14$ & $ 0.03\pm0.09$ & $ 0.25\pm0.06$ & $-0.44\pm 0.16$ \\
502 & 563,259 & $-8.65\pm0.04$ & $-0.62\pm0.07$ & $-0.05\pm0.06$ & $ 0.22\pm0.04$ & $-0.59\pm 0.09$ \\
539 & 426,270 & $-9.37\pm0.03$ & $-0.74\pm0.08$ & $ 0.11\pm0.06$ & $ 0.50\pm0.04$ & $-0.83\pm 0.09$ \\
865 & 496,331 & $-8.81\pm0.04$ & $-0.50\pm0.14$ & $ 0.05\pm0.10$ & $ 0.38\pm0.07$ & $-0.54\pm 0.16$ \\
894 & 362,338 & $-8.97\pm0.03$ & $-0.88\pm0.11$ & $-0.16\pm0.10$ & $ 0.63\pm0.07$ & $-0.77\pm 0.13$ \\
975 & 492,349 & $-8.76\pm0.05$ & $-0.44\pm0.11$ & $ 0.24\pm0.07$ & $ 0.58\pm0.04$ & $-0.62\pm 0.12$ \\
1094 & 520,364 & $-10.18\pm0.02$ & $-0.78\pm0.04$ & $-0.05\pm0.04$ & $ 0.51\pm0.03$ & $-0.75\pm 0.05$ \\
1112 & 303,367 & $-9.04\pm0.03$ & $-1.05\pm0.06$ & $-0.26\pm0.06$ & $-0.26\pm0.05$ & $-0.87\pm 0.08$ \\
1236 & 472,382 & $-9.04\pm0.04$ & $-0.34\pm0.23$ & $ 0.26\pm0.12$ & $ 0.72\pm0.07$ & $-0.54\pm 0.25$ \\
1443 & 292,408 & $-9.27\pm0.03$ & $-0.80\pm0.06$ & $-0.04\pm0.05$ & $ 0.43\pm0.03$ & $-0.78\pm 0.07$ \\
1448 & 370,407 & $-9.28\pm0.03$ & $-1.06\pm0.14$ & $ 0.26\pm0.13$ & $ 0.18\pm0.09$ & $-1.27\pm 0.18$ \\
1499 & 302,415 & $-9.37\pm0.03$ & $-0.93\pm0.06$ & $-0.15\pm0.06$ & $-0.31\pm0.05$ & $-0.83\pm 0.08$ \\
1688 & 222,442 & $-9.08\pm0.03$ & $-0.74\pm0.09$ & $ 0.08\pm0.07$ & $ 0.64\pm0.04$ & $-0.81\pm 0.10$ \\
1805 & 435,459 & $-10.31\pm0.02$ & $-0.82\pm0.04$ & $-0.01\pm0.03$ & $ 0.12\pm0.02$ & $-0.82\pm 0.04$ \\
1950 & 257,479 & $-9.23\pm0.03$ & $-0.39\pm0.09$ & $ 0.32\pm0.06$ & $ 0.96\pm0.03$ & $-0.64\pm 0.10$ \\
2228 & 358,529 & $-9.07\pm0.03$ & $-0.56\pm0.12$ & $-0.07\pm0.08$ & $ 0.52\pm0.06$ & $-0.51\pm 0.13$ \\
2284 & 536,536 & $-8.52\pm0.04$ & $-0.90\pm0.09$ & $-0.22\pm0.08$ & $-0.23\pm0.07$ & $-0.75\pm 0.10$ \\
2848 & 389,758 & $-8.78\pm0.03$ & $-0.39\pm0.11$ & $ 0.17\pm0.07$ & $ 0.53\pm0.04$ & $-0.52\pm 0.12$ \\
\enddata
\end{deluxetable}

\begin{deluxetable}{lrcc}
\tablecaption{\label{tab:clages}Cluster ages}
\tablecomments{
$A_B^i$ is the internal reddening for the cluster, determined from the
reddening map. age$_{\rm no corr}$ and age$_{\rm corr}$ are
the S-sequence ages in Myr without and with correction for internal reddening.
}
\tablehead{
  Object & $A_B^i$ & age$_{\rm no corr}$ (Myr) & age$_{\rm corr}$ (Myr) \\
}
\startdata
99 & 0.00 & 61 & 61 \\
502 & 0.53 & 31 & 27 \\
539 & 0.46 & 12 & 10 \\
865 & 0.81 & 42 & 35 \\
894 & 0.93 & 12 & 8 \\
975 & 0.70 & 38 & 33 \\
1094 & 0.58 & 15 & 12 \\
1112 & 0.58 & 4 & 3 \\
1236 & 0.64 & 53 & 46 \\
1443 & 0.59 & 13 & 10 \\
1448 & 0.84 & 1 & 1 \\
1499 & 0.61 & 8 & 6 \\
1688 & 1.05 & 13 & 10 \\
1805 & 0.52 & 11 & 9 \\
1950 & 1.03 & 39 & 31 \\
2228 & 0.87 & 41 & 33 \\
2284 & 0.20 & 12 & 11 \\
2848 & 0.00 & 51 & 51 \\
\enddata
\end{deluxetable}

\begin{deluxetable}{lrrr}
\tablecaption{\label{tab:ftab}
Data for CMD fits
}
\tablecomments{$N$ is the number of stars in the observed CMD}
\tablehead{
  Field & $A_B$ & $\chi^2$ & $N$ \\
}
\startdata
All & 0.0  &  4.4 & 2567\\
All & 0.0--1.0 &  1.0 & 2567\\
All & 0.0--2.0 &  3.6 & 2567\\
$0<r<100$ pixels & 0.0--1.0 &  2.3 & 560 \\
$0<r<100$ pixels & 0.0--2.0 &  3.2 & 560 \\
$100<r<300$ pixels & 0.0--1.0 &  1.2 & 1539 \\
$100<r<300$ pixels & 0.0--2.0 &  1.6 & 1539 \\
$0<r<100$ pixels & 0.0--1.0 &  2.3 & 560 \\
$100<r<150$ pixels & 0.0--1.0 &  2.1 & 527 \\
$150<r<200$ pixels & 0.0--1.0 &  1.3 & 499 \\
$200<r<250$ pixels & 0.0--1.0 &  0.9 & 316 \\
\enddata
\end{deluxetable}

\end{document}